\documentclass	
  [superscriptaddress,amsmath,aps,
    a4paper, 
    preprint, 
    floatfix
  ]
  {revtex4-1}
\usepackage[utf8x]{inputenc}
\usepackage[T1]{fontenc}
\usepackage{amsmath}
\usepackage{hyperref}
\usepackage{graphicx}
\usepackage{textcomp}
\usepackage{amssymb}
\usepackage[usenames,dvipsnames]{color}

\newcommand{\Ht}{\ensuremath{{^3\text{He}}}}
\newcommand{\Hf}{\ensuremath{{^4\text{He}}}}

\newcommand{\tcb}{\textcolor{black}}
\newcommand{\tcr}{\textcolor{black}}

\newcommand{\RED}{\textcolor{black}}

\newcommand{\REDCOLOR}{\textcolor{black}}

\begin{document}


\title{Tricritical Casimir forces and order parameter profiles in wetting films of $\Ht$ -$\Hf$ mixtures}
\author{N. Farahmand Bafi}
\email{nimabafi@is.mpg.de}
\affiliation{Max-Planck-Institut f{\"u}r Intelligente Systeme, Heisenbergstr. 3, D-70569 Stuttgart, Germany}
\affiliation{Institut f{\"u}r Theoretische Physik IV, Universit{\"a}t Stuttgart,Pfaffenwaldring 57, D-70569 Stuttgart, Germany}

\author{A. Macio\l ek}
\email{maciolek@is.mpg.de }
\affiliation{Institute of Physical Chemistry, Polish Academy of Sciences, Kasprzaka 44/52, PL-01-224 Warsaw, Poland}
\affiliation{Max-Planck-Institut f{\"u}r Intelligente Systeme, Heisenbergstr. 3, D-70569 Stuttgart, Germany}

\author{S. Dietrich}
\email{dietrich@is.mpg.de }
\affiliation{Max-Planck-Institut f{\"u}r Intelligente Systeme, Heisenbergstr. 3, D-70569 Stuttgart, Germany}
\affiliation{Institut f{\"u}r Theoretische Physik IV, Universit{\"a}t Stuttgart,Pfaffenwaldring 57, D-70569 Stuttgart, Germany}


\begin{abstract}

Tricritical Casimir forces in $\Ht$ -$\Hf$ wetting films are
studied, within mean field theory, in therms of
a suitable lattice gas model for binary liquid mixtures with
short--ranged
surface
fields.
The proposed model takes into account the continuous rotational symmetry O(2) of
the superfluid degrees of freedom associated with $\Hf$
and
it
allows, inter alia,
for the occurrence of a vapor phase. As a result,
the model
facilitates the formation
of wetting films,
which provides a
strengthened
theoretical framework to describe
available experimental data
for 
tricritical Casimir forces
acting
in $\Ht$ -$\Hf$ wetting films.
\end{abstract}


\maketitle

\section{Introduction}
\label{sec:intro}

Concerning
fluid
wetting films
near a critical point~\cite{generalphasediagram},
experimental studies
have provided
convincing evidence for
a long-ranged
effective interaction
emerging between
the planar
solid
surface and the parallel
fluid
interface forming the film~\cite{Chanhe4,indirect,Ganshin,Fukuto,Rafai,Mukhopadhyay,Mukhopadhyay2}.
Such fluid-mediated
and fluctuation induced
interactions were discussed
first
by Fisher and de Gennes \cite{Fischer-deGennes-1978} on the basis of  finite-size
scaling ~\cite{FSS,FSS2}  for 
critical
binary
liquid
mixtures. 
They are known
as {\it critical Casimir forces}
(CCFs) in  analogy
with
the well-known 
Casimir forces
in quantum
electrodynamics~\cite{casimiroriginal,krech}.
In  wetting films of
a classical binary liquid mixture,
within its bulk phase diagram
the
CCF arises
near
the
critical
end point
of the liquid
mixture,
at which
the line of critical points
of the liquid-liquid demixing transitions
encounters
the liquid-vapor coexistence
surface~\cite{generalphasediagram,nightingale}.
They  originate from the restriction
and modification of the
critical fluctuations of the  composition of the mixture   imposed  
on one side
by the solid substrate and
on the other side by
the
emerging
liquid--vapor interface.
The CCF acts
by moving
the liquid-vapor interface
and,
together
with the omnipresent background dispersion forces 
and
gravity, it
determines the equilibrium thickness $\ell$ of the wetting films \cite{Fukuto,Rafai,Mukhopadhyay,Mukhopadhyay2}.
The dependence of $\ell$  on temperature $T$ provides an indirect
measurement of
CCF~\cite{generalphasediagram,nightingale}. This approach also allows one
to probe the universal properties of
the
CCF
encoded in its scaling function~\cite{generalphasediagram}.
By  varying the undersaturation
of the vapor phase
one can  tune 
the film thickness and thus determine the  scaling behavior  of
the
CCF as function of $T$ and
$\ell$~\cite{generalphasediagram,prl66,1886}.
The
shape of such a
universal scaling function   depends on the bulk
universality class   of the confined fluid, and on the surface universality classes of 
the 
two
confining
boundaries~\cite{Diehl}.
The latter are related to the boundary conditions 
(BCs)~\cite{Diehl,krech,danchev} 
imposed by the surfaces on the  order parameter (OP)
associated with
the underlying
second-order phase transition~\cite{danchev}. In general,  the scaling function 
of
CCFs
is negative
(attractive CCFs)
for  symmetric BCs and positive
(repulsive CCFs)
for non-symmetric ones.  
Classical binary liquid mixtures near their demixing transition belong
to the $3d$ Ising universality class.
The
surfaces confining them belong to the so-called normal transition \cite{Diehl}, which is characterized by
a strong effective surface field acting on the deviation of the concentration from its critical value serving as the OP. The surface field describes the preference of the
surface for one of the two species forming the binary liquid mixture.
Since the two surfaces typically exhibit opposite preferences
wetting films of classical binary
liquid
mixtures are
often
characterized by opposing surface fields ($(+,-)$ BCs), which
results in
repulsive CCFs~\cite{Fukuto,Rafai,Mukhopadhyay,Mukhopadhyay2}.

In wetting films  of $\Hf$~\cite{Chanhe4}, the CCF originates
from the confined critical
fluctuations associated with the
continuous
superfluid
phase
transition
along the so--called $\lambda$--line. 
Similarly as for the  classical binary liquid mixtures,
here
the CCF emerges  near
that
critical
end point where the $\lambda$--line
encounters
the line of first--order liquid--vapor phase transitions of $\Hf$.

Capacitance measurements  of the equilibrium thickness of $\Hf$
wetting films  have provided
strong
evidence
for
an attractive CCF~\cite{Chanhe4,Ganshin} in
quantitative
agreement with
the theoretical predictions ~\cite{generalphasediagram,1886}
for the
corresponding
bulk $XY$ universality class
with \textit{symmetric}
Dirichlet-Dirichlet BCs (O,O), which
correspond
to the vanishing
of the
superfluid OP
with $O(2)$ symmetry both
at the surface of the substrate and at the liquid-vapor interface.
The scaling
function
of
this
CCF
has, to a certain extent,
been determined analytically ~\cite{generalphasediagram,1886,Zandi-Rudnick-Kardar-2004,Maciolek-Gambassi-Dietrich-2007,Zandi-Shackel-Rudnick-Kardar-Chayes-2007}
and by using
Monte Carlo simulations
~\cite{Vasilyev-Gambassi-Maciolek-Dietrich-2007,Vasilyev-Gambassi-Maciolek-Dietrich-2009,Dantchev-Krech-2004,Hutch-2007,hasenbusch-2009,hasenbusch-2010}.
Their results
are in
excellent
agreement with the experimental
data.

\begin{figure}
\includegraphics[width=100mm]{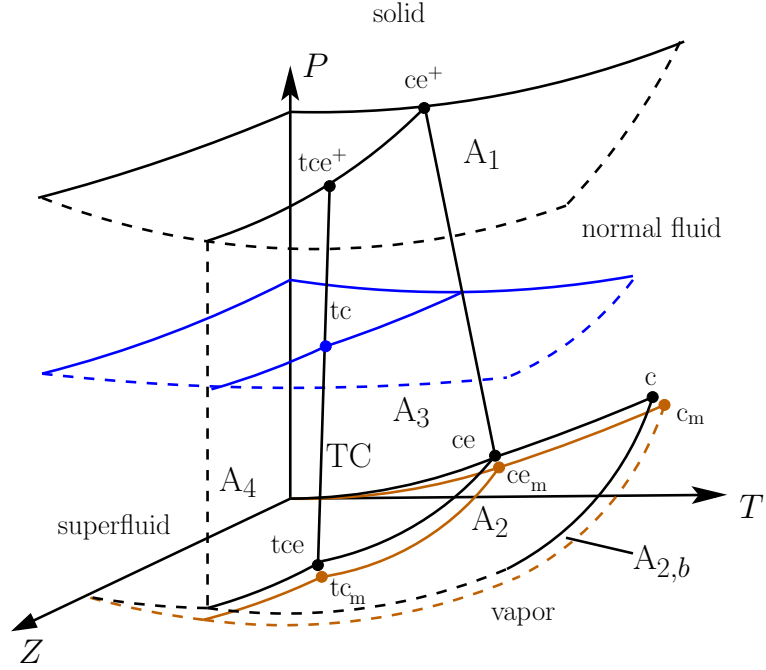}
\centering
\centering
\caption{\linespread{0.8}
\footnotesize{Schematic
bulk
phase diagram of $\Ht$ -$\Hf$ mixtures (black curves and surfaces)
and two specific
surfaces
(blue and brown)
in the $(T,Z,P)$
space,
where $P$ is the pressure and
$Z=\exp(\mu_3/T)$
is the fugacity of $\Ht$, with $\mu_3$
as
the chemical potential of $\Ht$
atoms~\cite{generalphasediagram}.
A$_1$ shows the surface of first-order
solid-liquid phase transitions, whereas A$_2$ is the
surface of first-order
vapor-liquid
phase transitions.
The phase transitions between
the normal fluid and
the
superfluid phase are either of
second or of first order,
which
are shown by the surfaces A$_3$ and A$_4$,
respectively.
The surfaces A$_3$ and A$_1$ intersect
along a line
ce$^+$-tce$^+$
of critical
end points.
The surfaces A$_3$ and A$_4$ are separated by
a
line
tce$^+$-tce
of tricritical points TC. This line
meets A$_1$ and A$_2$ at the tricritical end points
tce$^+$ and tce, respectively.
The surfaces A$_3$ and
A$_2$
intersect
along a line
ce-tce
of critical
end points.
The surface A$_2$ terminates at
a
line of critical points, starting from c
in
the plane
$Z=0$.
The phase diagram in
the plane
$Z=0$ corresponds to that of pure $\Hf$.
The dashed lines indicate that the corresponding surface continues.
On
the blue surface
the total density
is constant,
which corresponds to the situation
studied
in
Refs.~\cite{maciole-dietrich-2006,Maciolek-Gambassi-Dietrich-2007}.
The
brown
surface A$_{2,b}$ lies
in the vapor phase
slightly below
the
liquid-vapor coexistence surface A$_2$.
Although the thermodynamic fields along the thermodynamic paths
taken in the experiment
in Ref.~\cite{indirect} have been tuned to their values
at the liquid-vapor coexistence surface,
due to gravity
the actual measurements have been carried out for thermodynamic states
which
lie on
a surface
resembling
the brown one.
At the thermodynamic states on the brown
surface,
in addition to the stable vapor phase,
there
are
metastable
liquid phases. These metastable liquid phases
undergo transitions similar to
the
liquid--liquid phase transitions
tied to A$_2$.
Therefore, for
each point
tce, ce, and c, there is a metastable counter part
$\text{tc}_{\text{m}}$, $\text{ce}_{\text{m}}$, and $\text{c}_{\text{m}}$,
respectively,
on the brown surface.}
}
\label{3d_no_path}
\end{figure}

\begin{figure}
\includegraphics[width=85mm]{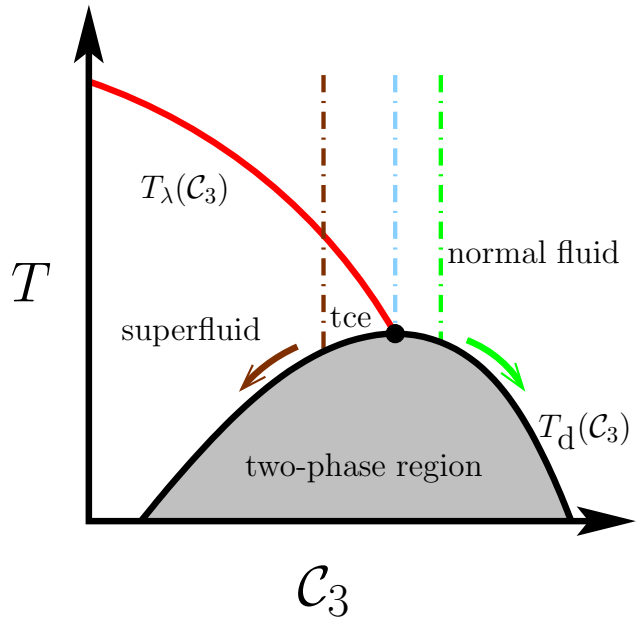}
\centering
\centering
\caption{\linespread{0.8}
\footnotesize{Liquid--liquid
bulk
phase transitions at coexistence with the vapor phase for $\Ht$ -$\Hf$ mixtures and
the thermodynamic paths
taken
in the experiments
reported
in Ref.~\cite{indirect}.
The black
curves denote
the first-order phase
transitions between the normal fluid
phase
and the superfluid
phase, which
terminate
at the
tricritical
end point
tce.
The red curve shows the
second--order
$\lambda$-transitions between
the normal fluid
phase
and the superfluid
phase.
The dashed dotted lines
indicate three distinct
thermodynamic paths corresponding to
three fixed values of the concentration
$\mathcal{C}_3=X_3/(X_3+X_4)$
(see, c.f., Eq.~(\ref{definition-x3_x4}))
of the $\Ht$ atoms
as done experimentally.
$X_3$ and $X_4$
are
the bulk number
densities
of $\Ht$ and $\Hf$, respectively.
Upon decreasing the temperature,
the bulk liquid undergoes a
first--order
phase separation at
some demixing temperature
$T_{\text{d}}(\mathcal{C}_3)$.
Upon further decrease
of the temperature
the thermodynamic paths follow
that
branch of the coexistence curve,
which they hit (see the
the brown and green
arrows).
}
}
\label{path_he43}
\end{figure}

\begin{figure}
\includegraphics[width=110mm]{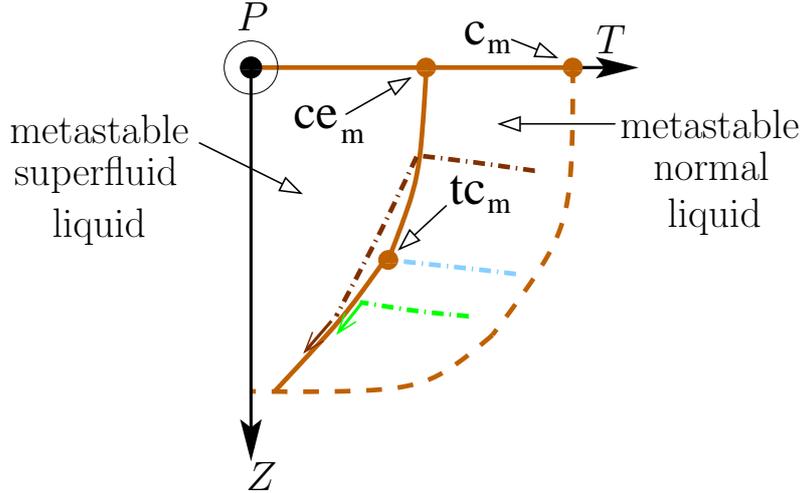}
\centering
\centering
\caption{\linespread{0.8}
\footnotesize{Projection
of the brown surface
A$_{2,b}$
in Fig.~\ref{3d_no_path}
(which lies in the vapor phase)
onto
the $(P,T)$ plane.
The solid and dashed brown lines are the projections of the
corresponding ones for A$_{2,b}$. The dashed--dotted lines are the
projections of
the
thermodynamic paths
shown in
Fig.~\ref{path_he43}
and which lie
on the brown surface.
Upon lowering $T$ the brown dashed--dotted line
first crosses the full line $T_\lambda(\mathcal{C}_3)$
in Fig.~\ref{path_he43}, continues through the superfluid phase, and then
encounters the two--phase region.
This
sketch is based
on our numerical results (see
Subsec.~\ref{subsec:he-wetting})
in the vicinity of
$\text{tc}_\text{m}$.
}
}
\label{path_on_offset_plane}
\end{figure}

Similar measurements \cite{indirect} for  wetting films of
$\Ht$ -$\Hf$
mixtures
performed  near the tricritical
end point,
at which the
line of tricritical points
encounters
the sheet of first--order liquid--vapor phase transitions
(see the phase diagram of
$\Ht$ -$\Hf$ mixtures
in Fig.~\ref{3d_no_path}),
revealed a
repulsive
tricritical
Casimir force (TCF).
In turn this points towards
non-symmetric BCs for the superfluid OP, which 
is
surprising
because
in this system
there are no surface fields
which
couple
to the superfluid OP.
However, there is a subtle physical mechanism which can create
(+,O) and thus non-symmetric BCs.
As argued in  Ref.~\cite{indirect}, the  $\Ht$ isotope is lighter than $\Hf$
and thus
experiences a larger
zero-point motion.
Hence it
occupies a larger volume than $\Hf$. As a result, $\Ht$ atoms are effectively
expelled from the
rigid
solid substrate
and
tend to gather at the
soft
liquid-vapor interface.
This leads to an effective attraction of
$\Hf$
atoms to the solid substrate so that a $\Hf$-rich layer forms near the substrate-liquid interface,
which
due to the increased $\Hf$ concentration
may become superfluid
at temperatures
already above the line of
onset of superfluidity in the bulk~\cite{Romagnan-et:1978}.
Thus the two interfaces impose a nontrivial concentration profile
across the film,
which in turn couples to the superfluid OP.
Explicit  calculations \cite{maciole-dietrich-2006,maciolek-gambassi-dietrich}
within the vectorized Blume-Emery-Griffiths (VBEG) model of helium
mixtures~\cite{beg,2dvbeg2,2dvbeg1}
have  demonstrated that the concentration profile indeed induces indirectly
non-symmetric
BCs for the
superfluid OP.
A
semi-quantitative
agreement with the experimental data given in Ref.~\cite{indirect}
has been found for the
TCF,
computed
by
assuming
a
symmetry-breaking $(+)$
BC
at the
substrate-liquid interface and
a
Dirichlet (O)
BC
at the liquid-vapor interface.
However, the VBEG model  employed in Refs.~\cite{maciole-dietrich-2006,maciolek-gambassi-dietrich}  does not incorporate the vapor phase
and hence cannot exhibit wetting films.
In these studies the confinement of the liquid between the substrate and the liquid-vapor interface has
been modeled by a slab geometry with
the boundaries introduced by fiat,
mimicking
the actual
self-consistent formation of wetting films and thus
differing
from the actual
experimental
setup.
This difference is
borne out
in Fig.~\ref{3d_no_path}.
Therein the
surface
of
constant
total density $D(P,T,Z)=const.$ is shown in blue.
The analyses
in Refs.~\cite{maciole-dietrich-2006,Maciolek-Gambassi-Dietrich-2007}
have been carried out within such a surface,
whereas
the experiment
in Ref.~\cite{indirect}
has
been
carried out
along the surface of liquid-vapor
coexistence. 
Note that,
although
the thermodynamic
states, for
which the measurements  have  been 
performed,
correspond to the liquid-vapor coexistence surface
(surface A$_2$
in Fig.~\ref{3d_no_path}),
due to
gravity  the actual thermodynamic paths lie 
on a surface, which is located
slightly
in the vapor phase 
(brown surface in Fig.~\ref{3d_no_path}).
Figures~\ref{path_he43}~and~\ref{path_on_offset_plane} show these thermodynamic paths.

In order
\REDCOLOR{to}
pave the way for providing
a more realistic description
of
the
experimental
setup reported
in Ref.~\cite{indirect}, 
recently
we have extended the VBEG model
such that
the vapor phase
is incorporated
into the phase diagram~\cite{bulk-phase-diagram}.
We have found that
allowing for the corresponding
vacancies
in the lattice model
leads
to
a rich phase behavior in the bulk
with complex phase diagrams of various topologies.
We were able to determine
that
range of interaction parameters  for which the bulk phase diagram
resembles
the one
observed experimentally for $\Ht$ -$\Hf$ mixtures, i.e.,
for which first--order
demixing
ends via a tricritical point
at the $\lambda$-line of second--order superfluid transitions~\cite{bulk-phase-diagram}.
In the present
study,
we use this model
in order
to describe  wetting
of a solid substrate by
$\Ht$ -$\Hf$ mixtures.
We
analyze
the behavior of
the
wetting films along the thermodynamic paths
corresponding to the ones
in the experiment~\cite{indirect}.
This will allow us to compare the
variation
of the wetting film  thickness
with the experimental data shown in Fig.~\ref{Indirect-Fig} of Ref.~\cite{indirect} (see Sec.~\ref{sec:lw}),
which
is
not possible within the 
approach used in Refs.~\cite{maciole-dietrich-2006,Maciolek-Gambassi-Dietrich-2007}.
Finally, we aim
at extracting
the TCF contribution to the effective
force
between the solid substrate and
the
emerging
liquid-vapor
interface. We shall
compare its  scaling
function
with
that
extracted from
the
experimental
data in Ref.~\cite{indirect} and
the one
calculated using the
simple
slab geometry
employed
in Refs.~\cite{maciole-dietrich-2006,Maciolek-Gambassi-Dietrich-2007}.
We
study
our model
in spatial dimension $d=3$ within mean field theory
which, up to
logarithmic
corrections, captures
the universal behavior  of the TCF near the tricritical point of  $\Hf$ -$\Hf$ mixtures.
However, this
approximation
is
insufficient near
the critical points of the
second--order
$\lambda$-transition, because
for the  tricritical phenomena the upper critical dimension
is
$d^*=3$, whereas 
for the critical ones 
it is
$d^*=4$.

Our paper is organized as follows.
In
Sec.~\ref{sec:model}
we introduce the model and in
Subsec.~\ref{subsec:mfa} we carry out a
mean field approximation
to it.
In
Subsec.~\ref{subsec:bpd}
we discuss a procedure for
finding
that
range of values of interaction
constants of the model
for which
it
exhibits a
phase
diagram similar to that of
actual
$\Ht$ -$\Hf$ mixtures.
We continue in Sec.~\ref{sec:lw} with studying the wetting films  for
short--ranged
surface fields.
Next, we
calculate
TCFs
and their scaling functions
and
compare our results with
those for
the
slab geometry
by applying a suitable slab approximation to the present case.
In
Sec. IV
we conclude with a summary.
Appendix A contains important technical details.

\section{The model}
\label{sec:model}

In order to model
$\Ht$ -$\Hf$ mixtures in
the presence of a
solid, two-dimensional
surface, we consider a three--dimensional
(d\,=\,3)
simple cubic lattice formed by $L$ layers of two--dimensional $N\times N$ lattices with
lattice spacing
$a$.
In the following all lengths are measured in units of $a$, which is equivalent to consider these lengths to be dimensionless together with setting $a=1$.
In each layer, all $\mathcal{N}:=N^2$ lattice sites are identical.
The different lattice sites are
label by $\{i\,|\,i=1,...,L \mathcal{N}\}$. Alternatively,
one can use the index $l$, labeling the layer
number,
and
the index $v_l$, referring to lattice sites within
the $l^\text{th}$ layer.
The lattice sites
$\{i\,|\,i=1,...,L \mathcal{N}\}=\{(l,v_l)\,|\,l=0,...,L-1;v_l=1,..., \mathcal{N}\}$
are occupied by either $\Ht$
or $\Hf$ atoms or they
are unoccupied. We
consider nearest-neighbor interactions with
the Hamiltonian
\begin{equation}
\label{main-Hamiltonian}
  \begin{split}
   \mathcal{H} & = -J_{44}N_{44}-J_{33}N_{33}-J_{34}N_{34}\\
     & \quad -(\mu_{4}+f_4(l))N_{4} -(\mu_{3}+f_3(l))N_{3} -J_{s}\tilde{N}_{44}\text{,}
   \end{split}
\end{equation}
where $N_{\text{m} \text{n}}$, with $\text{m},\text{n}\in \{3, 4\}$,
denotes the number of pairs of
nearest neighbors
of species
$^\text{m}\text{He}$ and $^\text{n}\text{He}$
on the lattice
sites.
$N_{\text{m}}$
denotes the number of
$^\text{m}\text{He}$
atoms
and $-J_{s}\tilde{N}_{44}$
is
the sum of the interaction
energies
between the superfluid degrees of freedom
$\Theta_{i}$ and $\Theta_{j}$
associated with the nearest--neighbor pairs $\langle i,j\rangle$ of $\Hf$ with
$J_s$ as the
corresponding
interaction strength
(see, c.f., Eq.~(\ref{Nvector})).
The
effective interactions between
pairs of helium isotopes
are represented by
$ J_{33}$, $ J_{44}$, and $ J_{34}$.
The
three
effective pair potentials between the
two types of
isotopes are not
identical
due
to their distinct statistics and the slight
differences
in
their
electronic states.
The
surface fields, which represent
the effective interaction between the surface and
the $\Hf$ and $\Ht$
atoms,
are denoted as $f_4(l)$ and $f_3(l)$, respectively.
In general these surface fields
depend
on the distance $l$ from the surface, which is located at
$l=0$, and vanish for large $\ell$.
The chemical potential of species
$^\text{m}\text{He}$ is denoted as $\mu_{\text{m}}$.
(The Hamiltonian in Eq.~(\ref{main-Hamiltonian}))
with $J_s=0$ describes a classical binary liquid mixture
of species m and n.)

In order to proceed, we associate
an occupation variable $s_{i}$
with each lattice site $\{i\}$, which can take the
three values $ +1$, $-1$, or $0$,
where $+1$ denotes that
the lattice site is occupied by
$\Hf$, $-1$ denotes that the lattice site is occupied by
$\Ht$,
and $0$ denotes that the lattice site is unoccupied.

$N_{\text{m}}$ and $N_{\text{m} \text{n}}$ can be expressed in terms of $\{s_i\}$ as follows:

\begin{equation}
\label{occupation-operator}
 \begin{split}
 &N_{4}=\frac{1}{2} \sum_{i} s_{i} (s_{i}+1)\equiv \sum_{i} p_i\text{,}\\
 &N_{3}=\frac{1}{2} \sum_{i} s_{i} (s_{i}-1)\text{,}\\
 &N_{44}=\frac{1}{4} \sum_{<i,j>} (s_{i} (s_{i}+1)s_{j} (s_{j}+1))\equiv \sum_{<i,j>} p_i p_j\text{,}\\
 &N_{33}=\frac{1}{4} \sum_{<i,j>} (s_{i} (s_{i}-1)s_{j} (s_{j}-1))\text{,}\\
 &N_{34}=\frac{1}{4} \sum_{<i,j>} (s_{i} (s_{i}+1)s_{j} (s_{j}-1)+ s_{i} (s_{i}-1)s_{j} (s_{j}+1) )\text{,}
 \end{split}
\end{equation}
where $\sum \limits_{<i,j>}$ denotes the sum over nearest neighbors. Using the above definitions one obtains
\begin{equation}
\label{hamiltonian-summation}
  \begin{split}
   \mathcal{H} & = -K\sum_{<i,j>}s_{i}s_{j}-J\sum_{<i,j>}q_{i}q_{j}-C\sum_{<i,j>}(s_{i}q_{j}+q_{i}s_{j})\\
     & \quad -\mu_{-}\sum_{i}s_{i}-\mu_{+}\sum_{i}q_{i}  -\sum_{i}f_-(l)s_{i}-\sum_{i}f_+(l)q_{i}\\
     &   \quad -J_{s}\sum_{<i,j>}p_{i}p_{j}\cos(\Theta_{i}-\Theta_{j})        \text{,}
  \end{split}
\end{equation}
where 
\begin{equation}
\label{Nvector}
      \sum_{<i,j>}p_{i}p_{j}\cos(\Theta_{i}-\Theta_{j})= \tilde{N}_{44}= \sum_{<i,j>}p_{i}p_{j}\begin{pmatrix}\cos\Theta_{i}\\ \sin\Theta_{i}\end{pmatrix}\cdot\begin{pmatrix}\cos\Theta_{j}\\ \sin\Theta_{j} \end{pmatrix},
\end{equation}
%
%
%
and
\begin{equation}
\label{coupling}
 \begin{split}
 &q_{i}=s_{i}^2\text{,}\\
 &p_{i}= \frac{1}{2} s_{i}(s_{i}+1)\text{,}\\
 &K=\frac{1}{4}(J_{44}+J_{33}-2J_{34})\text{,}\\
 &J=\frac{1}{4}(J_{44}+J_{33}+2J_{34})\text{,}\\
 &C=\frac{1}{4}(J_{44}-J_{33})\text{,}\\
 &\mu_{-}=\frac{1}{2}(\mu_{4}-\mu_{3})\text{,}\\
 &\mu_{+}=\frac{1}{2}(\mu_{4}+\mu_{3})\text{,}\\
 &f_+(l)=\frac{1}{2}(f_4(l)+f_3(l))\text{,}\\
 &f_-(l)=\frac{1}{2}(f_4(l)-f_3(l))\text{.}
 \end{split}
\end{equation}
$\Theta_{i}\in [0,2\pi]$ represents the superfluid degree of freedom at the lattice site i, provided it is occupied by $\Hf$.


\subsection{Mean field approximation}
\label{subsec:mfa}

In this section we
carry out a
mean field approximation
for
the present model (for details of the
calculations
see Appendix A).
The
symmetry of the problem
implies
that
all
statistical quantities exhibit the same mean values for all
lattice sites within a
layer, in particular
the same mean field generated by their neighborhood.
Therefore all
quantities depend only
on the distance $l$ of a layer from the surface.
(Note
that $l$ is an integer which not only represents the position of the layer but also marks the corresponding
layer.)
We define the following
dimensionless OPs:
\begin{equation}
\label{definition of order parameters}
 \begin{split}
   & X_l:=\langle s_{(l,v_l)}\rangle\text{,}\\
   & D_l:=\langle q_{(l,v_l)}\rangle\text{,}\\
   & M_l^2:=\langle p_{(l,v_l)}\sin\Theta_{(l,v_l)}\rangle^2+\langle p_{(l,v_l)}\cos\Theta_{(l,v_l)}\rangle^2 \text{,}
 \end{split}
\end{equation}
which are coupled by the following self-consistent
equations:
\begin{equation}
\label{eqX}
X_l=\frac{-W_l+R_l I_0(\beta J_s \tilde{M_l})}{1+W_l+R_l I_0(\beta J_s\tilde{M_l})}\text{,}
\end{equation}
\begin{equation}
\label{eqD}
D_l=\frac{W_l+R_l I_0(\beta J_s \tilde{M_l})}{1+W_l+R_l I_0(\beta J_s \tilde{M_l})}\text{,}
\end{equation}
and 
\begin{equation}
\label{eqM-easy}
M_l=\frac{R_l I_1(\beta J_s \tilde{M_l})}{1+W_l+R_l I_0(\beta J_s \tilde{M_l})}\text{,}
\end{equation}
where
$\beta=1/T$ with $T$
as
temperature times $k_\text{B}$,
$I_{0}(\beta J_s \tilde{M_l})$ and $I_{1}(\beta J_s \tilde{M_l})$
are modified Bessel functions, and

\begin{equation}
\tilde{M_l}=(1-\delta_{l,0})M_{l-1}+4 M_{l}+M_{l+1}\text{.}
\end{equation}
The dimensionless
functions $W_l$ and $R_l$ depend on the following
set of parameters:
\\
$(X_l,D_l;\mu_{-},\mu_{+},f_+(l),f_-(l),T)$.
They
are given by
\begin{equation}
\label{EqW}
\begin{split}
W_l(X_l,D_l;\mu_{-},\mu_{+},f_+(l),f_-(l),T)=\exp\Big[\beta\{&(J-C) (D_{l-1}(1-\delta_{l,0})+4 D_{l}+D_{l+1})\\
						    & +(C-K)(X_{l-1}(1-\delta_{l,0})+4 X_{l}+X_{l+1})\\
						    &+ \mu_{+}+f_{+}(l)- \mu_{-}-f_{-}(l)\}\Big]
\end{split}
\end{equation}
and
\begin{equation}
\label{EqR}
\begin{split}
R_l(X_l,D_l;\mu_{-},\mu_{+},f_+(l),f_-(l),T)=\exp\Big[\beta\{&(J+C) (D_{l-1}(1-\delta_{l,0})+4 D_{l}+D_{l+1})\\
						    &+(C+K)(X_{l-1}(1-\delta_{l,0})+4 X_{l}+X_{l+1})\\
						    &+ \mu_{+}+f_{+}(l)+\mu_{-}+f_{-}(l)\}\Big].
\end{split}
\end{equation}

Accordingly, the
equilibrium free energy per
number of lattice sites in a single layer
is given by 
\begin{equation}
\label{free-energy-simple}
\begin{split}
 \phi/\mathcal{N}=& \sum_{l=0}^{L-1} \Big [\frac{K}{2} X_l(4X_l+X_{l+1}+X_{l-1}(1-\delta_{l,0}))\\
		  & \quad +\frac{J}{2} D_l(4D_l+D_{l+1}+D_{l-1}(1-\delta_{l,0}))\\
		  & \quad + \frac{C}{2} X_l(4D_l+D_{l+1}+D_{l-1}(1-\delta_{l,0}))\\
		  & \quad + \frac{C}{2} D_l(4X_l+X_{l+1}+X_{l-1}(1-\delta_{l,0}))\\
		  & \quad + \frac{J_s}{2} M_l(4M_l+M_{l+1}+M_{l-1}(1-\delta_{l,0}))\\
		  & \quad +(1/\beta)\ln (1-D_l) \Big ].
\end{split}
\end{equation}
Within the grand--canonical ensemble the pressure is $P=-\phi/V$, where
here the volume
\tcr{is}
$V=L\mathcal{N}a$, with $a=1$.
The
functional form of the expressions
\tcr{for}
the
chemical potentials are obtained by solving
Eqs.~(\ref{eqX})~and~(\ref{eqD})
for them
\tcr{(see Appendix A):}
\begin{equation}
\label{deltaPLUS}
 \begin{split}
     \mu_{+}=&\, \frac{T}{2}\ln (D_l^2-X_l^2)-T\ln2-T\ln(1-D_l)-\frac{T}{2}\ln(I_0(\beta J_s \tilde{M_l}))\\
                                  & -J(D_{l-1}(1-\delta_{l,0})+4D_{l}+D_{l+1})-C(X_{l-1}(1-\delta_{l,0})+4X_{l}+X_{l+1})-f_{+}(l),
 \end{split}
\end{equation}
and
\begin{equation}
\label{deltaMINUS}
\begin{split}
\mu_{-}=&\, \frac{T}{2}\ln\frac{D_l+X_l}{D_l-X_l}-\frac{T}{2}\ln(I_0(\beta J_s \tilde{M_l}))\\
			     & -C(D_{l-1}(1-\delta_{l,0})+4D_{l}+D_{l+1})-K(X_{l-1}(1-\delta_{l,0})+4X_{l}+X_{l+1})-f_{-}(l).
\end{split}
\end{equation}
Finally,
one can express the magnetization $M_l$ in terms of $X_l$
and $D_l$ by using Eqs.~(\ref{eqX})~to~(\ref{eqM-easy}):
\begin{equation}
\label{mEQUILIBRIUM}
\frac{X_l+D_l}{2}=\frac{M_l I_0(\beta J_s \tilde{M_l})}{I_1(\beta J_s \tilde{M_l})}.
\end{equation}

According to the definition of the OPs
in Eq.~(\ref{definition of order parameters})
and
by
using
Eqs.~(\ref{occupation-operator})~and~(\ref{breaking-sum})
one can
express
the number
densities
of
species $\Hf$ and $\Ht$
in the $l^{\text{th}}$ layer
as 
\begin{equation}
\label{definition-x3_x4}
\begin{split}
&X_{4,l}=\frac{\langle N_{4,l}\rangle}{\mathcal{N}}=\langle p_{l}\rangle=\frac{1}{2}\left \langle s_{l} (s_{l}+1)\right \rangle=\frac{D_l+X_l}{2},\\
&X_{3,l}=\frac{\langle N_{3,l}\rangle}{\mathcal{N}}=\frac{1}{2}\left \langle s_{l} (s_{l}-1)\right \rangle=\frac{D_l-X_l}{2},
\end{split}
\end{equation}
so that $D_l=X_{4,l}+X_{3,l}=\langle s_{l}^2 \rangle$ and
$X_l=X_{4,l}-X_{3,l}=\langle s_{l}  \rangle$,
where $s_l\equiv s_{(l,v_l)}$ is 
the occupation variable of a single lattice site within the $l^{\text{th}}$ layer;
its thermal average
is independent of $v_l$ (see Appendix A).
Accordingly, the concentration of the two species
in
the $l^{\text{th}}$ layer is
given
by
$\mathcal{C}_{4,l}\equiv\frac{X_{4,l}}{X_{4,l}+X_{3,l}}=\frac{D_l+X_l}{2D_l}$
and
$\mathcal{C}_{3,l}\equiv\frac{X_{3,l}}{X_{4,l}+X_{3,l}}=\frac{D_l-X_l}{2D_l}$.

In order to study wetting films at
given values of $(T,\mu_+,\mu_-)$,
one has to
solve the set of equations
given by
Eqs.~(\ref{deltaPLUS})~-~(\ref{mEQUILIBRIUM})
for the set of OPs
$\{(X_l,D_l,M_l)\,|\,l=0,...,L-1\}$.
Since
Eqs.~(\ref{eqX})~-~(\ref{eqM-easy}) cannot
be solved analytically,
we did so numerically
by using
the
GSL library~\cite{gsl-cite}.
Since
for the last layer $l=L-1$
Eqs.~(\ref{deltaPLUS})~to~(\ref{mEQUILIBRIUM})
request
OP values
at $l=L$,
one has to assign values to $(X_{L+1},D_{L+1},M_{L+1})$.
If the system size $L$ is
sufficiently
large
one expects that far away from the surface
the OP profiles
attain
their bulk values.
This implies
$(X_{L},D_{L},M_{L})=(X_{\text{bulk}},D_{\text{bulk}},M_{\text{bulk}})$.
The
system size $L$
can be considered to be
large enough
if the OP profiles
$(X_l,D_l,M_l)$ remain
de facto
unchanged upon increasing $L$ (which mimics a semi-infinite system).
The minimization procedure, which
leads
to Eqs.~(\ref{deltaPLUS})~-~(\ref{mEQUILIBRIUM})
does not
involve
the second derivative of $\phi$ with respect to
the trial density matrix $\rho_l$ (see Appendix A).
Therefore, depending on the initial profile
$\{(X_l,D_l,M_l)\,|\,l=0,...,L-1\}$, with which one starts the
iteration algorithm,
the solution of Eqs.~(\ref{deltaPLUS})~-~(\ref{mEQUILIBRIUM}) might
correspond to
a local minimum, a  local maximum,
or a saddle point.


\subsection{Bulk phase diagram}
\label{subsec:bpd}

Since the realization of the experimental paths in Ref.~\cite{indirect}
requires the knowledge of the bulk phase diagram,
first one has to find the set of coupling constants, for which
the model exhibits a
phase
diagram similar to that of
actual
$\Ht$ -$\Hf$ mixtures.
This
issue has been addressed
in Ref.~\cite{bulk-phase-diagram}.
Here we summarize
those
results of these studies
which are relevant for the present analysis.

\begin{figure}
\includegraphics[width=76mm]{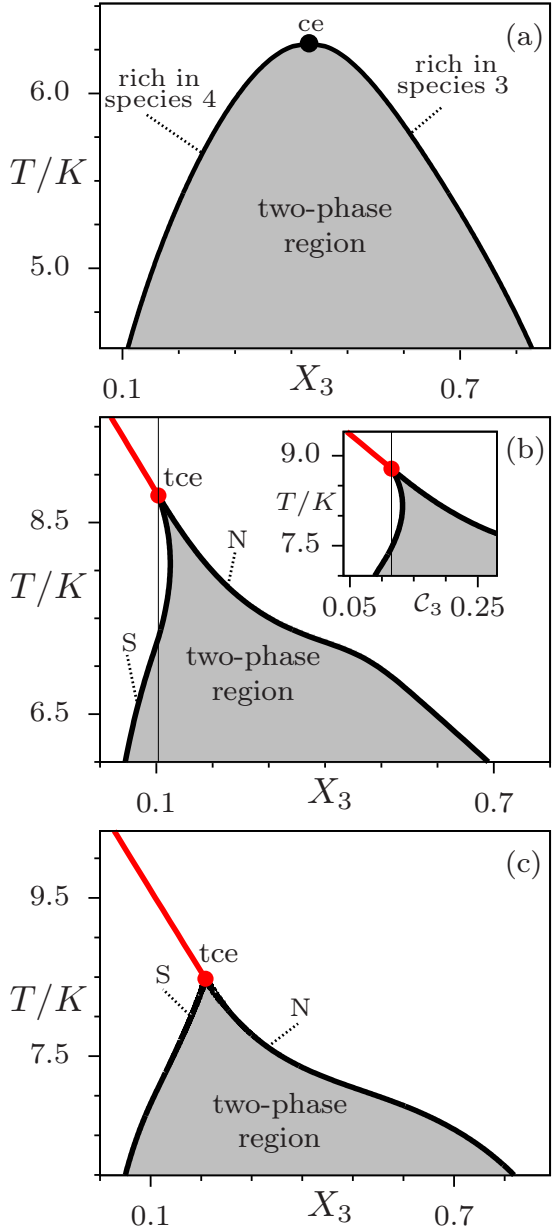}
\centering
\centering
\caption{\linespread{0.8}
\footnotesize{Liquid--liquid demixing
phase transitions
in the bulk
at coexistence with the vapor phase (the vapor phase is not shown here)
in the $(X_3$, $T)$ plane, with
$X_3=\langle N_3\rangle/(L\mathcal{N})=D-X$
for
(a) $(C/K, J/K, J_s/K)=(1, 5.714, 0)$, (b) $(C/K, J/K, J_s/K)=(1, 5.714, 3.674)$, and (c) $(C/K, J/K, J_s/K)=(1, 9.107, 3.701)$.
The inset of panel (b) shows the same phase diagram
in the
$(T,\mathcal{C}_3)$
plane, where $\mathcal{C}_3=X_3/D$
denotes the concentration
of $\Ht$.
The phase diagrams in (a) and (b)
have been discussed in detail in Ref.~\cite{bulk-phase-diagram}.
(Note
that in Ref.~\cite{bulk-phase-diagram} the coupling constants
are rescaled by a factor of 6 and the total
number of lattice sites are
denoted as
$\mathcal{N}$,
whereas
here
the total number of lattice sites is
given by $L\mathcal{N}$.) 
In (b) and (c)
the black curves denote the
binodals
of the
first--order
phase
transitions
between the normal fluid (N) and the superfluid (S).
The
lines
of
first--order
phase transitions in (a)
terminate
at the critical
end point
ce
with
$T_{\text{ce}}/K=6.286$, whereas 
in (b) and (c) the
lines
of
first--order
phase transitions
terminate
at
a tricritical end point
tce.
In (b) and (c)
the red curve
denotes the $\lambda$-line
of
second--order
phase transitions between the normal fluid and the superfluid.
The temperature of the tricritical
end point
in panels (b) and (c) are
$T_{\text{tce}}/K=8.782$ and $T_{\text{tce}}/K=8.47974$,} respectively.
In (b) the thin vertical line indicates $X_3=X_3^{\text{tce}}$,
whereas in the inset of this figure the thin vertical line
indicates $\mathcal{C}_3=\mathcal{C}_3^{\text{tce}}$.
The short dotted strokes indicate the character
(S, N, rich in species $4$, or rich in species $3$) of the corresponding binodal.
}
\label{different_topologies}
\end{figure}

\begin{figure}
\includegraphics[width=95mm]{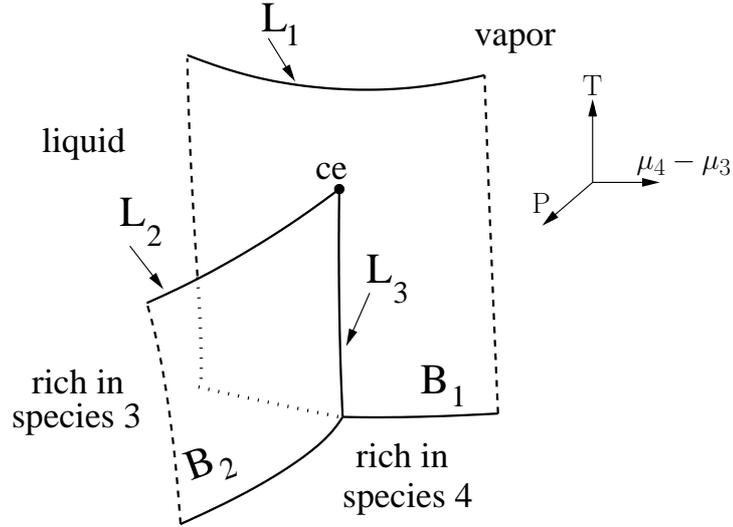}
\centering
\centering
\caption{\linespread{0.8}
\footnotesize{Fluid parts of the phase diagram of
a classical binary liquid mixture
in the $(T,P,\mu_4-\mu_3)$
space
(schematic diagram).
B$_1$
is the surface of first--order liquid--vapor phase transitions,
whereas
B$_2$
is the surface of first--order liquid--liquid demixing transitions,
between phases rich in either species 3 or 4.
The surface
B$_1$
terminates
at a line L$_1$ of critical points.
L$_2$ denotes the line of critical points of the 
liquid--liquid demixing transitions, which ends at the surface
B$_1$
at the critical end point ce. The surfaces
B$_1$ and B$_2$
intersect along a line L$_3$ of triple points.
The dashed curves
indicate
that the
corresponding
surfaces
continue.
The
demixing two--phase region in terms
of temperature
and the
number density
$X_3$ of species 3 for
the
liquid phases coexisting along
L$_3$ are shown in Fig.~\ref{different_topologies}(a);
the vapor phase
is not shown there.
}}
\label{binary_fluid_schematic}
\end{figure}

\begin{figure}
\includegraphics[width=85mm]{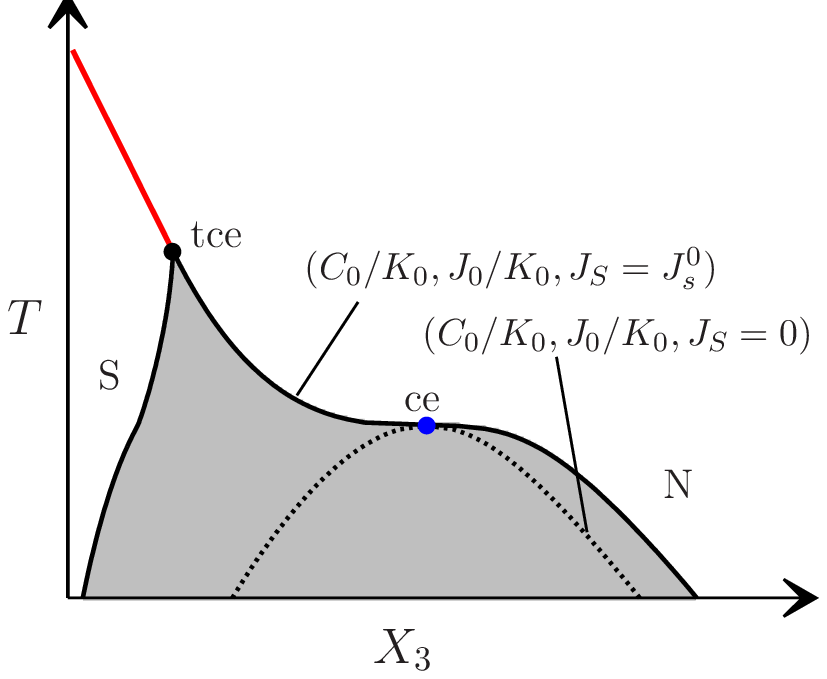}
\centering
\centering
\caption{\linespread{0.8}
\footnotesize{Schematic representation of the transformation of 
the
bulk
phase diagram of
a
classical binary
liquid
mixture
for
fixed values of $(C_0/K_0, J_0/K_0)$ and $J_s=0$
(the dotted curve)
into that of $\Ht$-$\Hf$ mixtures
with $J_s=J_s^0\neq0$ (solid curves).
In a first
step,
for suitable,
fixed values of $(C_0/K_0, J_0/K_0)$ and $J_s=0$
one
has
the phase diagram of a classical binary mixture
with ce as in
Fig~\ref{different_topologies}(a).
In a second
step,
one has to find
a nonzero value of $J_s=J_s^0$
(which produces the superfluid phase) such,
that
the critical end point ce of the phase diagram
for $(C_0/K_0, J_0/K_0, J_s=0)$,
is in thermodynamic coexistence
with a superfluid phase.
For this new set of coupling constants
$(C_0/K_0, J_0/K_0, J_s>J_s^0)$,
the phase diagram
with $(C_0/K_0, J_0/K_0, J_s=0$)
lies in the two phase region of the phase diagram with
$(C_0/K_0, J_0/K_0, J_s>J_s^0)$.
}}
\label{topology_condition}
\end{figure}

Taking
the OPs
to be
independent of $l$ and
omitting
the surface fields,
i.e., $f_{+}(l)=f_{-}(l)=0$, Eqs.~(\ref{eqX})~-~(\ref{eqM-easy}),
and Eqs.~(\ref{deltaPLUS})~-~(\ref{mEQUILIBRIUM}),
together with the expression for the equilibrium free energy given by
Eq.~(\ref{free-energy-simple}),
render the bulk phase diagram of the system
as
studied in Ref.~\cite{bulk-phase-diagram}.
It 
has been demonstrated in Ref.~\cite{bulk-phase-diagram}
(see also Ref.~\cite{vbeg3})
that
various
coupling constants lead to
diverse
topologies
of the phase diagram for
the bulk liquid--liquid demixing transitions.
The topologies discussed in Ref.~\cite{bulk-phase-diagram}
range  from
the phase diagram of a classical binary mixture
(Figs.~\ref{different_topologies}(a)~and~\ref{binary_fluid_schematic})
to
a phase diagram which to a large extent resembles the actual one
of
\REDCOLOR{$\Ht$ -$\Hf$ mixtures}
(Fig.~\ref{different_topologies}~(b)).
By extending the corresponding discussion in Ref.~\cite{bulk-phase-diagram}
one can study how, within
the present model,
for a suitable value of $J_s$ the bulk
phase diagram of
a
classical binary mixture with specific values of
$(C_0/K_0, J_0/K_0)$ and
for
$J_s=0$
(dotted
curve in Fig.~\ref{topology_condition})
transforms
into
that of the $\Ht$-$\Hf$ mixture.
Figure~\ref{topology_condition}
illustrates
schematically this transformation.
One has to
find and to adopt
a nonzero value of $J_s=J_s^0$ such that
the critical
end point
ce of the phase diagram
for $(C_0/K_0, J_0/K_0, J_s=0)$
is in
thermodynamic coexistence with
a
superfluid phase.
This locates the critical
end point
ce on the right
shoulder of the transformed
phase diagram. Thus
for $J_s>J_s^0$, the initial  phase diagram
for $(C_0/K_0, J_0/K_0, J_s=0)$ (including its critical
end point
ce),
lies in the
two--phase
region of the phase diagram for $(C_0/K_0, J_0/K_0, J_s>J_s^0)$~\cite{bell}.
Although the phase diagram in Fig.~\ref{different_topologies} (b)
satisfies the above condition and captures
the main
features of the bulk phase diagram
of $\Ht$-$\Hf$ mixtures, its shape near the tricritical
end point
tce
differs from the experimental one (see Fig.~\ref{path_he43}).
In particular,
in the phase diagram in Fig.~\ref{different_topologies} (b),
upon
lowering
the temperature below $T_\text{tce}$
along the path $X_3=X_3^\text{tce}$,
the
model
mixture does not enter the
two--phase
region, as
it is the case for
the actual
$\Ht$-$\Hf$
mixture.
Note that
the
experimental phase diagram
in Fig.~\ref{path_he43}
is drawn in the $(T,\mathcal{C}_3)$ plane.
(The
model
phase diagram
in the
same
$(T,\mathcal{C}_3)$ plane is
shown  in the inset of Fig.~\ref{different_topologies} (b))
Furthermore, although the condition $J_s>J_s^0$ places the critical
end point
ce of the
phase diagram with $(C_0/K_0, J_0/K_0, J_s=0)$
into the two--phase
region
of the phase diagram with $(C_0/K_0, J_0/K_0, J_s>J_s^0)$,
a certain
residual,
distorting
influence of this critical end point ce
on the wetting 
films may still
be present,
especially if ce lies near any of the two
binodals
of the demixing
transitions of the transformed
phase diagram
(solid black lines in Figs.~\ref{different_topologies}(b)~and~(c)).
In order to
address this issue,
after finding the necessary
conditions for the coupling parameters leading to
the desired topology,
we have modified the values of
$(C_0/K_0,J_0/K_0)$
with $J_s=J_s^0$
such, that the critical end point ce (which starts to
shift into metastablity for $J_s=J_s^0$)
moves deeply into the
two-phase
region of the
transformed
phase diagram.
These
considerations have led us to choosing the following
choice for the
coupling constants:
$(C/K, J/K, J_s/K)=(1, 9.10714, 3.70107)$. The corresponding
phase diagram is shown in Fig.~\ref{different_topologies} (c).

\section{Layering and wetting for short--ranged surface fields}
\label{sec:lw}

In this section we study
the
layering and  wetting
behavior~\cite{dietrich-wetting}
of the present model with
short--ranged
surface fields  
$f_+(l)=\tilde{f_+}\delta_{l,0}$ and $f_-(l)=\tilde{f_-}\delta_{l,0}$.
The field $f_+(l)$ describes  the enhancement of
the
fluid density near the wall, whereas $f_-(l)$
expresses
the preference of the wall for
$\Hf$ over $\Ht$.

Within the present model $\mu_{+}$ is the field conjugate to the number density order parameter $D_l$.
By changing $\mu_{+}$ from its value
$\mu_{+}^\text{co}(P,T)$ at
liquid-vapor coexistence
and
at a given
temperature $T$ and pressure $P$,
one
can drive the bulk system either towards the liquid phase ($\Delta \mu_+=\mu_+-\mu_+^{\text{co}}>0$)
or towards the vapor phase
($\Delta \mu_+<0$).
In order to realize the experimental conditions we
choose
$\Delta \mu_+<0$
such that the bulk system remains
thermodynamically
in the vapor phase.
With this constraint we determine the solution of
Eqs.~(\ref{deltaPLUS})~-~(\ref{mEQUILIBRIUM})
for set of the OPs
$\{(X_l,D_l,M_l)\,|\,l=0,...,L-1\}$.
We find that
the occurrence of wetting films as well as their thicknesses depend on the strength of the surface fields
$\tilde{f_+}$ and $\tilde{f_-}$.
Since along the experimental
paths
taken
in Ref.~\cite{indirect} the system is in the complete wetting regime,
we
choose such
values of the surface fields for which
complete wetting does occur. We refrain from exploring the
full variety of
scenarios for  wetting
transitions
which
can occur
within
the present model.

Based on the number density profile
$D_l$
one can define the film thickness as~\cite{dietrich-wetting}
\begin{equation}
\label{thicknessDefinition}
y(\mu_{-},\mu_{+},\tilde{f_+},\tilde{f_-},T)=\frac{\varrho}{D_{\text{m}}-D_{\text{b}}},
\end{equation}
where
$D_b$ is the bulk density of the vapor phase,
\begin{equation}
\label{excess}     
\varrho=\sum_{l=0}^{L-1} (D_l-D_{\text{b}})
\end{equation}
is the excess
adsorption,
and $D_\text{m}$ is the density of the metastable
liquid phase at the  thermodynamic state corresponding
to
the stable vapor phase.
Alternatively, one can define $y$
as the position of the
inflection point of the density profile
$D_l$
at the
emerging
liquid-vapor
interface.
The profile
$X_l=X_{4,l}-X_{3,l}$ indicates,
whether the
various
layers are
occupied mostly by
species of type $4$ (positive or large values of $X_l$) or by species of type $3$ (negative or small values).
A nonzero  magnetization profile
signals
that  the wetting film is  superfluid.
In the following
subsections
we present our results
for
$J_s=0$ (Subsec.~\ref{subsec:js_zero}),
which  corresponds to a classical  binary  mixture, and
$J_s\ne 0$ (Subsec.~\ref{subsec:he-wetting}),
which corresponds
to a $\Ht$-$\Hf$ mixture.
The former case shows how within the present model the strength of the
surface fields influences
the formation and the thickness of the wetting films, whereas
the latter case focuses on describing the
present,
experimentally
relevant situation.


\subsection{Layering and wetting for $J_s=0$}
\label{subsec:js_zero}
\begin{figure}
\includegraphics[width=115mm]{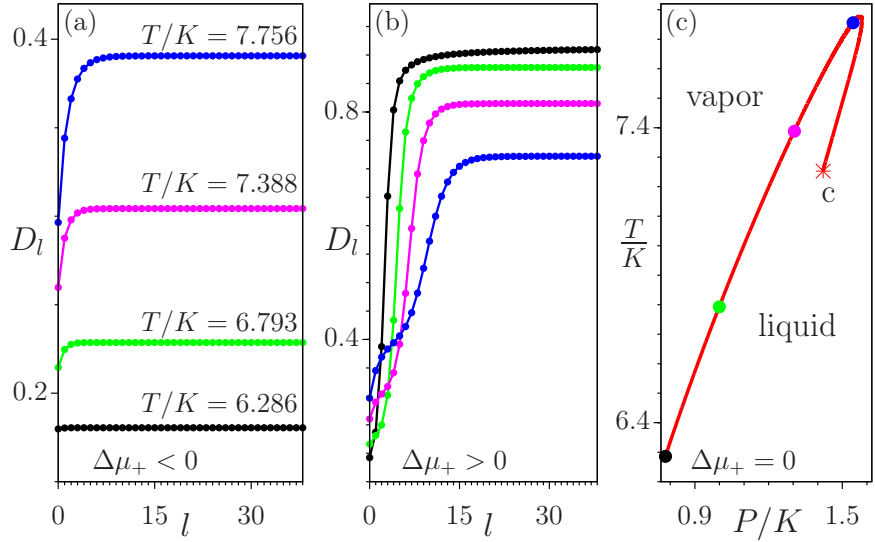}
\centering
\centering
\caption{\linespread{0.8}
\footnotesize{Number density profiles
for
weak surface fields
$(\tilde{f_+},\tilde{f_-})/K=(0.857,0)$
\tcr{and}
for several values of
\tcr{temperature $T$}
(same color code in all panels)
above $T_\text{ce}/K=6.286$
and at 
$X_3=X_3^{\text{ce}}=0.431$
\tcr{in}
the liquid
\tcr{phase, with
(a)
$\Delta \mu_{+}/K=-8.57\times 10^{-4}$ and (b)
$\Delta \mu_{+}/K=+8.57\times 10^{-4}$.
In (a) the bulk phase is the vapor phase whereas in (b) it is the
coexisting liquid phase with slight offsets.}
Panel (c) shows the liquid--vapor coexistence
\tcr{line}
(red curve)
in the $(P,T)$
\tcr{plane, emerging}
under the constraint $X_3=X_3^{\text{ce}}$
\tcr{in}
the liquid phase.
The colored
\tcr{dots}
in panel (c) indicate the thermodynamic
states with the same temperature values as in
\tcr{panels}
(a) and (b),
\tcr{which, unlike in}
these two panels, lie at liquid--vapor coexistence.
The star denotes the liquid--vapor critical point
$(P_\text{c}/K,T_\text{c}/K)=(1.422,7.230)$.
The thermodynamic paths in
\tcr{panels}
(a) and (b) follow the red curve
in panel (c) but with the corresponding offset
\tcr{values}
$\Delta \mu_{+}$.
In panel (a)
the corresponding thermodynamic states are:
$(T/K,D=X_4+X_3,X=X_4-X_3,M)=\{(6.286,0.181,-0.121,0),(6.793,0.229,-0.137,0),(7.388,0.304,-0.155,0),(7.756,0.391,-0.174,0)\}$.
\tcr{The vapor bulk phase in (a) is}
preferred
by the wall. Accordingly, there are no
\tcr{liquidlike}
wetting films.
In panel (a)
near $T_\text{c}$ critical adsorption
(see Fig.~8 in Ref.~\cite{dietrich-surface-films}) of the
preferred vapor phase
occurs, which is indicated by the increased
depth and range of the minimum in $D_l$.
\tcr{In panel (b), due to $\Delta \mu_{+}>0$}
the stable
bulk
phase is
\tcr{the}
liquid.
\tcr{Since the}
vapor phase is preferred by the wall drying films form there
upon increasing the temperature.
The corresponding thermodynamic states are:
$(T/K,D,X,M)=\{(6.286,0.913,0.050,0),(6.793,0.878,0.015,0),(7.388,0.815,-0.048,0),(7.756,0.722,-0.141,0)\}$.
The value of $\mu_-$ can be obtained from
\tcr{Eq.~(\ref{deltaMINUS})}
using the
values
of
$(T,D,X,M)$ for the corresponding thermodynamic states
as
provided above.
\tcr{Note that the nonmonotonic behavior of the red curve in (c) is caused
by the constraint $X_3=X_3^{\text{ce}}$.}
\tcr{For (c), in order to identify the vapor and liquid phases,
in addition to $T$ and $P$
also the
value of the chemical potential $\mu_-$ is required, which is not shown.}
}
}
\label{small_f_plus}
\end{figure}

\begin{figure}
\includegraphics[width=125mm]{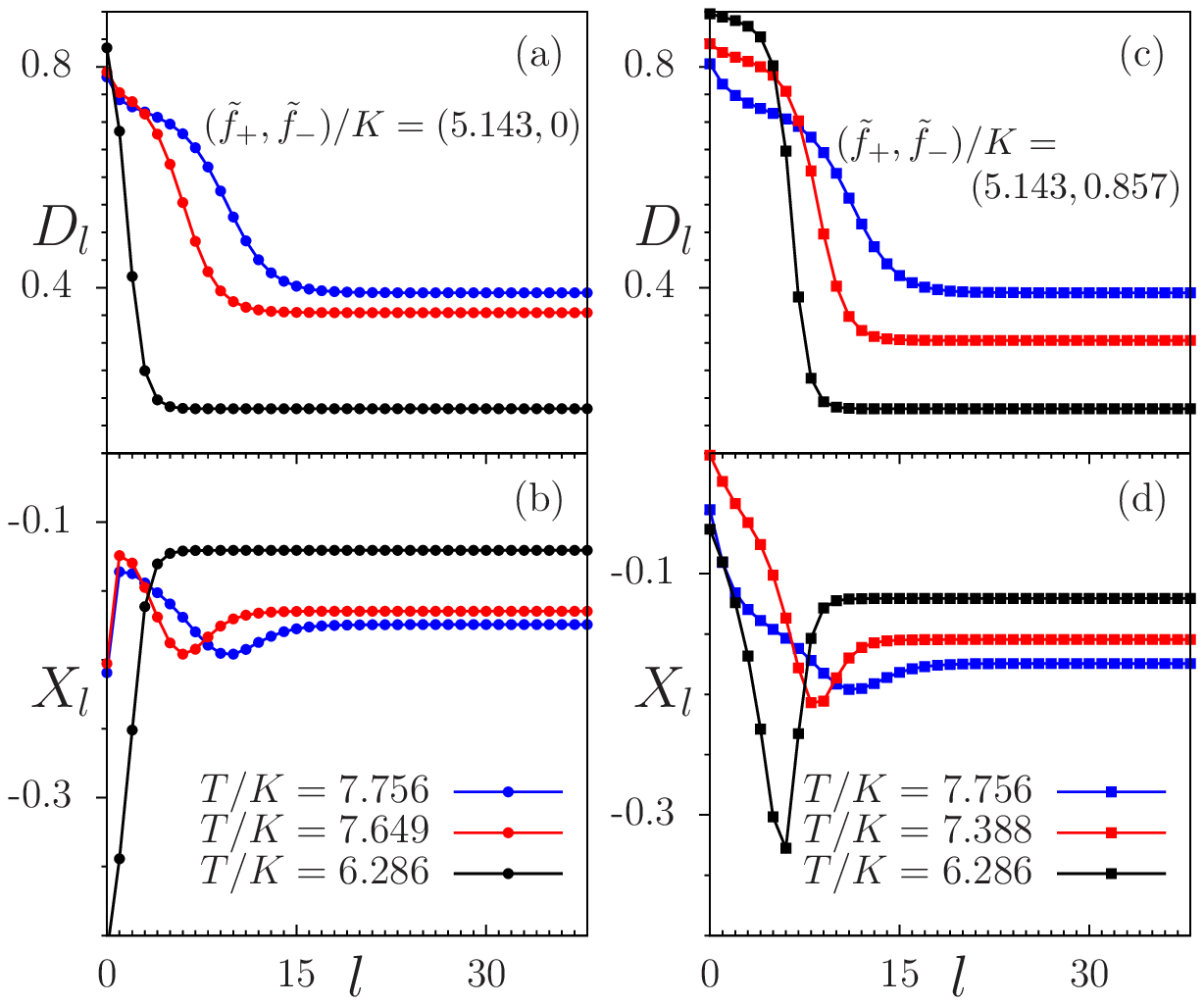}
\centering
\centering
\caption{\linespread{0.8}
\footnotesize{Order parameter profiles
$D_l$ and $X_l$ at $X_3=X_3^{\text{ce}}=0.431$ and
for
$\Delta \mu_{+}/K=-8.57\times 10^{-4}$
for several temperatures
and
for two sets of surfaces fields.
In panels
(a) and (b) the surface fields are $(\tilde{f_+},\tilde{f_-})/K=(5.143,0)$,
whereas panels (c) and (d)
correspond to
$(\tilde{f_+},\tilde{f_-})/K=(5.143,0.857)$.
The bulk phase is the vapor phase and the wall prefers
the liquid phase, giving rise to wetting films.
The positive value of $\tilde{f_-}$
not only results in the increase of the number density
$X_{4,l}$
of species $4$
and hence also
of
$X_l$
in the first layer (see panel (d)),
but also increases the total number density $D_l$ in the first layer (see panel (c)).
The stable thermodynamic states of the vapor phase in the bulk are
$(T/K,D,X,M)=\{(6.286,0.180,-0.121,0),(7.388,0.304,-0.155,0),(7.649,0.355,-0.165,0),(7.756,0.391,-0.174,0)\}$.
All three
\tcr{bulk}
states lie in the vapor phase
close to liquid--vapor coexistence
on the left
\tcr{side}
of the red line shown
Fig.~\ref{small_f_plus}(c).}
The value of $\mu_-$ can be obtained from
\tcb{Eq.~(\ref{deltaMINUS})}
using the
values
of
$(T,D,X,M)$ for the corresponding thermodynamic states
as
provided above.
}
\label{middle_f_plus-with-and-without-fminus}
\end{figure}

\begin{figure}
\includegraphics[width=85mm]{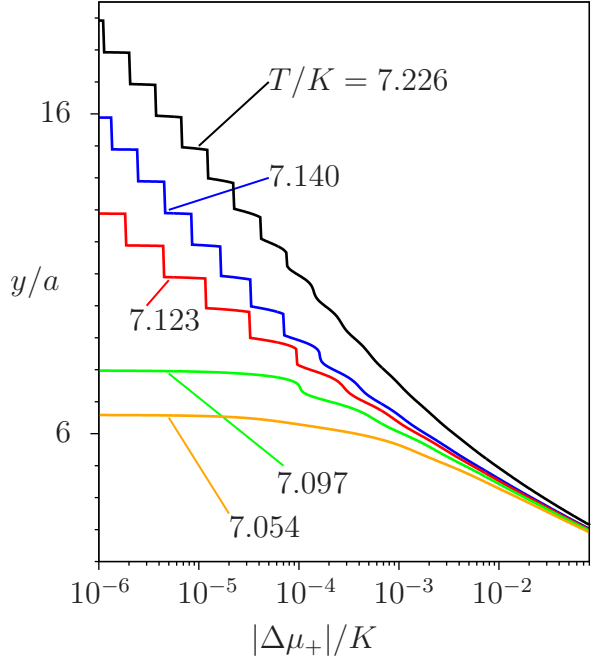}
\centering
\centering
\caption{\linespread{0.8}
\footnotesize{Film thickness $y/a$ versus
$|\Delta\mu_{+}|/K$
for
$(\tilde{f_+},\tilde{f_-})/K=(8.571,0)$
at
$X_3=X_3^{\text{ce}}$, and for
several
temperatures.
Upon approaching the liquid-vapor coexistence surface at low temperatures,
the film thickness increases smoothly
and reaches
a plateau.
The height of this plateau increases gradually by increasing $T$,
indicating $7.097<T_w/K<7.123$.
The jumps are due to
first--order
layering transitions
induced by the lattice model.
Above the roughening transition they are an artifact of mean field theory~\cite{dietrich-wetting}.
For $T>T_w$ one has
$y(\Delta\mu_{+}\to 0^-)\sim \kappa\text{ln}\frac{1}{|\Delta\mu_+|/K}$
with a slight increase of $\kappa(T)$ as a function of $T$.
}}
\label{changing-offset-normal}
\end{figure}

\begin{figure}
\includegraphics[width=85mm]{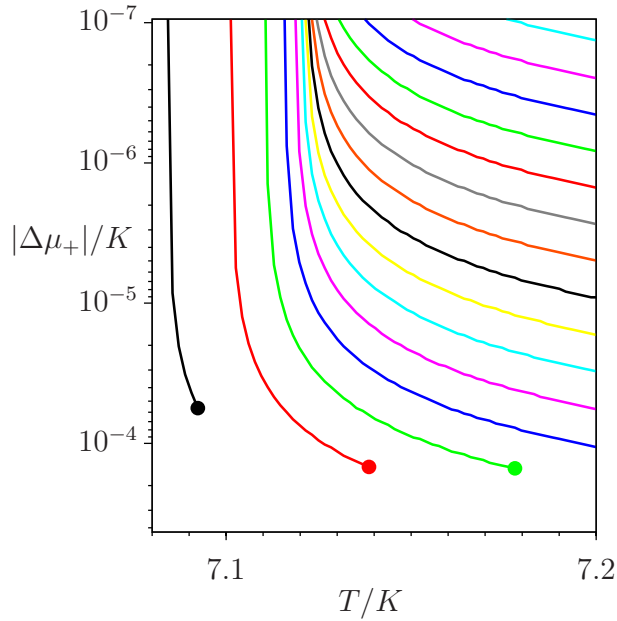}
\centering
\centering
\caption{\linespread{0.8}
\footnotesize{Layering transitions in the $(\mu_+,T)$ plane for
$(\tilde{f_+},\tilde{f_-})/K=(8.571,0)$.
Each line of
first--order layering
transition ends
at a critical point.
The color code in Fig.~\ref{layering} does not
carry a specific meaning.
The
lines are colored differently
so that it is easier to distinguish them.
}}
\label{layering}
\end{figure}

\begin{figure}
\includegraphics[width=95mm]{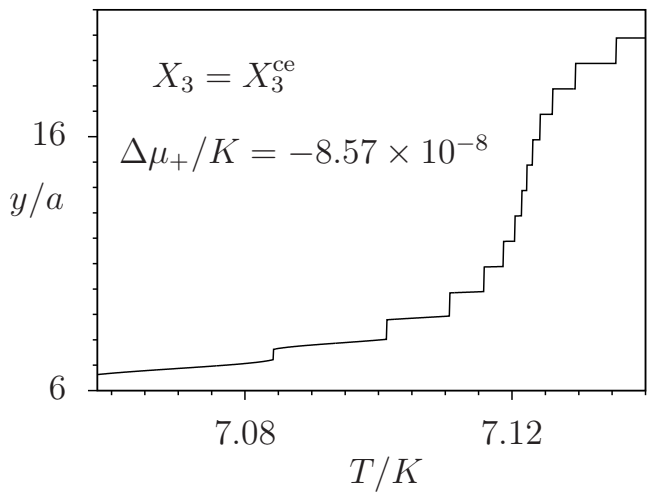}
\centering
\centering
\caption{\linespread{0.8}
\footnotesize{Equilibrium film thickness $y/a$
as a function of temperature for
$\Delta \mu_+/K=-8.57\times10^{-8}$ and
$X_3=X_3^{\text{ce}}$.
Since $\Delta\mu_+$ is nonzero, $y(T)$ does not diverge but attains a maximum
upon passing by $T_w$. This maximum diverges for $\Delta\mu_+\to0$.
The jumps are bunched together around $T/K\simeq 7.125$ and
spread--out
otherwise.
}}
\label{film_thickness_Normal_case}
\end{figure}

In this
subsection
we consider
a classical binary liquid mixture
of
species $3$ and $4$,
described by the Hamiltonian given
in
Eq.~(\ref{hamiltonian-summation}) 
with the coupling constants $(C/K, J/K, J_s/K)=(1, 5.714, 0)$.
The bulk phase
diagram
of
this system in the $(T,X_3)$ plane
is shown in Fig.~\ref{different_topologies}(a).
All figures in this subsection
(i.e., Figs.~\ref{small_f_plus}~-~\ref{film_thickness_Normal_case})
share the coupling constants $(C/K, J/K, J_s/K)=(1, 5.714, 0)$.
\RED{In Figs.~\ref{small_f_plus}(a)~and~\ref{middle_f_plus-with-and-without-fminus}~-~\ref{film_thickness_Normal_case}
the system size is $L=40$, whereas in Fig.~\ref{small_f_plus}(b)
it is $L=80$.}
We study how the
strengths
of the surface fields influence the formation of the wetting films for
thermodynamic states
with $T>T_\text{ce}$ and $\Delta \mu_{+}=\mu_{+}-\mu_{+}^\text{co}<0$,
i.e.,
corresponding to the vapor
being the bulk phase and the wetting phase being the mixed supercritical liquid phase.

We
start our discussion by taking $\tilde{f_-}=0$ and varying $\tilde{f_+}$.
We find that weak surface fields $\tilde{f_+}$ cannot stabilize high density layers near the surface, so that
the model does not exhibit wetting
by the mixed--liquid phase.
Instead,
for weak $\tilde{f}_+$ the wall
prefers the vapor phase so that upon approaching the liquid--vapor coexistence
from the liquid side (i.e., $\Delta\mu_+\to 0^+$)
a vapor film forms close to the wall corresponding to drying of
the interface between the wall and the mixed liquid.

Covering the case of weak surface
fields,
Fig.~\ref{small_f_plus}(a) shows the number density profiles
for
$(\tilde{f_+},\tilde{f_-})/K=(0.857,0)$
at
$\Delta \mu_{+}/K=(\mu_{+}-\mu_{+}^\text{co})/K=-8.57\times 10^{-4}$,
i.e., on the vapor side
for
several temperatures above the
$T_\text{ce}$ and at
fixed $X_3=X_3^{\text{ce}}$.
Figure.~\ref{small_f_plus}(b)
shows the number density profiles
for the same bulk system with the same surface fields but for
$\Delta \mu_{+}/K=(\mu_{+}-\mu_{+}^\text{co})/K=+8.57\times 10^{-4}$
so that the stable bulk phase
is liquid.
Since the wall prefers the vapor phase,
upon increasing
$T$
a
drying film forms at the
surface of the
solid substrate.

For larger values of $\tilde{f_+}$
(see Fig.~\ref{middle_f_plus-with-and-without-fminus}),
i.e., for
$(\tilde{f_+},\tilde{f_-})/K=(5.143,0)$,
at lower temperatures $T$
we find monotonically decaying density profiles
without shoulder formation
whereas  at higher temperatures the density
profiles
tend to
exhibit plateaus characteristic of wetting
(see Fig.~\ref{middle_f_plus-with-and-without-fminus}(a)).
Note that
in Figs.~\ref{middle_f_plus-with-and-without-fminus}(a)~and~(b)
the number density in the first layer
as part of the wetting film
decreases upon increasing $T$.
This is in accordance with the fact that
the density of the bulk liquid phase as the wetting phase
decreases upon heating, whereas the
bulk
vapor density increases.
The
profiles $X_l=X_{4,l}-X_{3,l}$ 
shown in red and blue
in Fig.~\ref{middle_f_plus-with-and-without-fminus}(b)
have  local minima at
$l=5$ and $l=9$, respectively.
These minima
occur
approximately
at the position of the
emerging
liquid-vapor interface
(see the corresponding curves
in panel (a)) and
indicate 
that species
of type $3$ preferentially
accumulate at
the liquid-vapor interface.
Figures~\ref{middle_f_plus-with-and-without-fminus}~(c)~and~(d)
show the OP
profiles for
$(\tilde{f_+},\tilde{f_-})/K=(5.143,0.857)$
and 
$\Delta \mu_{+}/K=-8.57\times 10^{-4}$
at several
values of the temperature. One can see that for 
positive values of $\tilde{f_-}$ both  $D_l$
and $X_l$ are enhanced in the first layer. This corresponds to the preferential adsorption 
of
species of type $4$
at the wall.

In order to see how the wetting films
grow
upon approaching the liquid-vapor coexistence surface, we
fix $T$ and
vary
$\Delta \mu_{+}$.
Figure~\ref{changing-offset-normal}
shows the film thickness
$y/a$ versus
$\Delta \mu_{+}$ for
$(\tilde{f_+},\tilde{f_-})/K=(8.571,0)$
and for
several temperatures;
$y$ is calculated
according to Eq.~(\ref{thicknessDefinition}).
For low temperatures, upon approaching the liquid-vapor coexistence surface
the film thickness increases smoothly and reaches a plateau.
This corresponds to
incomplete
wetting.
The height of this plateau increases gradually upon
increasing $T$
towards $7.097<T_w/K<7.123$,
which corresponds to
a critical wetting
transition
between
incomplete
and complete wetting~\cite{dietrich-wetting}.
The
corresponding
line of
wetting transitions lies on the surface of
the
liquid--vapor transitions
(B$_1$
in Fig.~\ref{binary_fluid_schematic})
between the critical end point ce and
and the line of critical points of the liquid--vapor transitions
(L$_1$ in Fig.~\ref{binary_fluid_schematic}).
Note that
Fig.~\ref{changing-offset-normal} provides a semi--logarithmic plot so that
the linear
growth of the film thickness on this scale confirms
the theoretically
expected
logarithmic growth of the film thickness
$y\sim \log (|\Delta \mu_+|/K)$
for
short--ranged
surface fields~\cite{dietrich-wetting}.
At higher temperatures
$T$ the film thickness does not
increase smoothly anymore
but rather exhibits jumps due to layering
transition.
Figure~\ref{layering}
shows the location of these
layering transitions in the $(\mu_+,T)$ plane for
$(\tilde{f_+},\tilde{f_-})/K=(8.571,0)$. If
a thermodynamic path passes through any of these lines
the film thickness
undergoes
a small jump of the size $l\simeq1$.
Each line of the layering
transitions
ends at a critical point.
Along thermodynamic paths, which pass by
these critical points the
jumps of the film thickness become rounded as
for
the green and red curves in Fig.~\ref{changing-offset-normal}.
The color code in
Fig.~\ref{layering} does not
carry a particular
meaning; the lines  are  colored
differently so that
it is easier to distinguish them.
The closer the system is to the liquid-vapor coexistence
surface, i.e.,
the smaller $\Delta \mu_+$ is,
the closer
are
the lines of layering transitions.
Figure~\ref{film_thickness_Normal_case}
shows how the increasing density of the layering transition lines affects
the film thickness while varying the temperature at fixed values of
$\Delta \mu_+/K=-8.57\times10^{-8}$ and $X_3=X_3^{\text{ce}}=0.431$.



\subsection{Layering and wetting for $J_s \ne 0$}
\label{subsec:he-wetting}

\begin{figure}
\includegraphics[width=105mm]{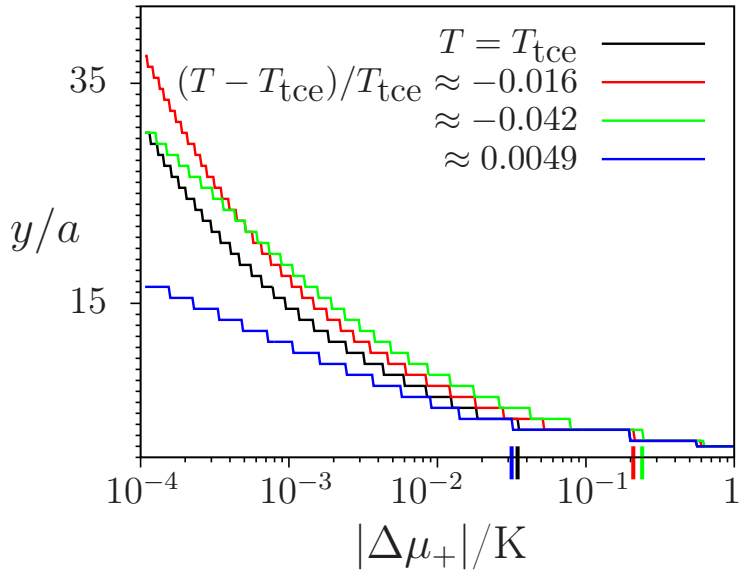}
\centering
\centering
\caption{\linespread{0.8}
\footnotesize{Equilibrium film thickness $y/a$ versus
$|\Delta \mu_{+}|/K$ for $(\tilde{f_+},\tilde{f_-})/K=(10.714,16.071)$
and for four
temperatures.
Unlike the situation in
Fig.~\ref{changing-offset-normal}
with $J_s=0$, the thickness
of
the
wetting films as a function of $|\Delta\mu_+|$
is a nonmonotonic function of  $T$.
The most rapid increase occurs at
$(T-T_{\text{tce}})/T_{\text{tce}}\approx -0.016$, whereas for lower and higher temperatures
the growth of the wetting film as a function of $\Delta\mu_+$
is slower.
Upon approaching the liquid-vapor coexistence surface,
the $\Hf$-rich
layers
within  the wetting
films
become superfluid.
At each temperature, the
continuous
surface transition to
superfluidity occurs for values of
the offset
$|\Delta\mu_+|$
smaller than the one indicated by 
the
corresponding
tick
on the abscissa
with the same color.
For $(T-T_\text{tce})/T_\text{tce}=-0.016$
and $(T-T_\text{tce})/T_\text{tce}=-0.042$, the number density of
$\Ht$
on the superfluid branch of the binodal (Fig.~\ref{different_topologies}(c))
is $X_3=0.201$ and $X_3=0.188$, respectively,
whereas for  $T\geq T_\text{tce}$
the number density of $\Ht$ is fixed
at
$X_3=X_3^{\text{tce}}=0.20845$.
}}
\label{changing_offset_X3_constant}
\end{figure}

\begin{figure}
\includegraphics[width=85mm]{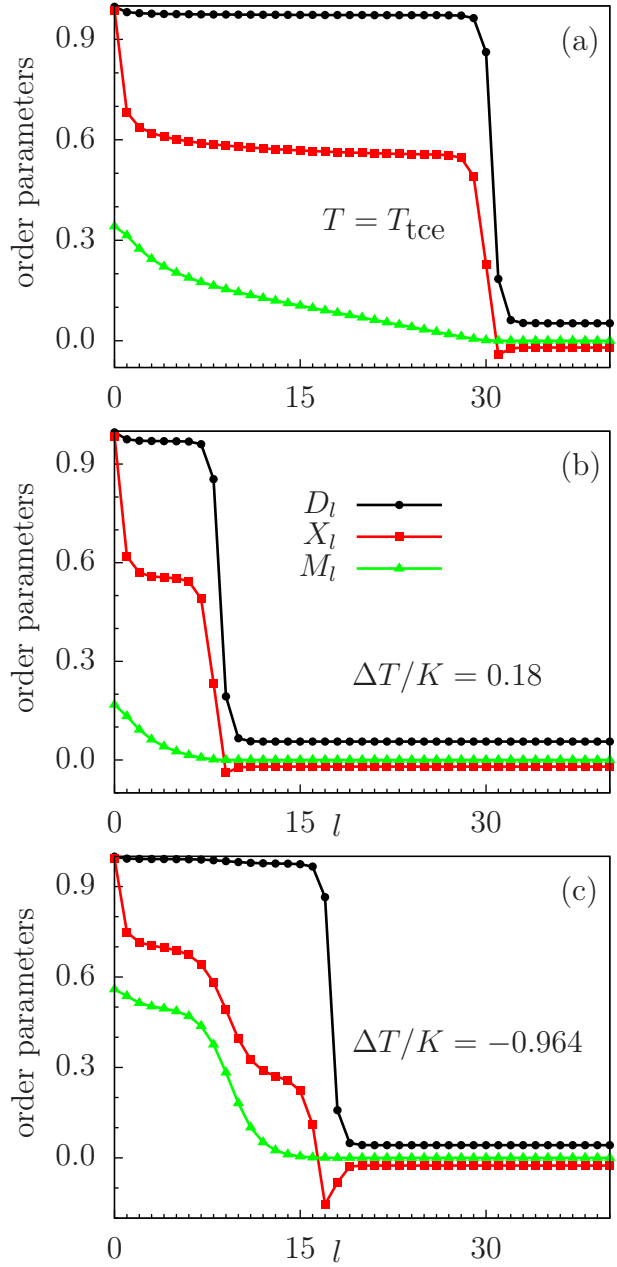}
\centering
\centering
\caption{\linespread{0.8}
\footnotesize{Order parameter profiles
$D_l$, $X_l$, and $M_l$
for
$(\tilde{f_+},\tilde{f_-})/K=(10.714,16.071)$
at
$\Delta \mu_+/K=-1.07\times 10^{-4}$ for
the bulk states
(a) $T=T_{\text{tce}}$, $X_3=X_3^{\text{tce}}$,
(b) $\Delta T/K=(T-T_{\text{tce}})/K=0.18$, $X_3=X_3^{\text{tce}}$, and 
(c) $\Delta T/K=(T-T_{\text{tce}})/K=-0.964$, $X_3=0.1461$
(which is on the superfluid branch of the binodal (Fig.~\ref{different_topologies}(c)).
For these bulk states,
in panels (a) - (c)
the stable vapor phase
(i.e., $l\to\infty$)
exhibits
the order parameters
$(D=X_4+X_3,X=X_4-X_3)=\{(0.0523,-0.0206),(0.0559,-0.0204),(0.0422,-0.0255)\}$,
respectively.
The bulk parameters of the system are those for Fig.~\ref{different_topologies}(c).
The keys for the OP profiles are the same for all panels.
The value of $\mu_-$ can be obtained from
\tcb{Eq.~(\ref{deltaMINUS})}
using the
values
of
$(T,D,X,M)$ for the corresponding thermodynamic states
as
provided above.
}}
\label{profiles_at_three-T}
\end{figure}

\begin{figure}
\includegraphics[width=100mm]{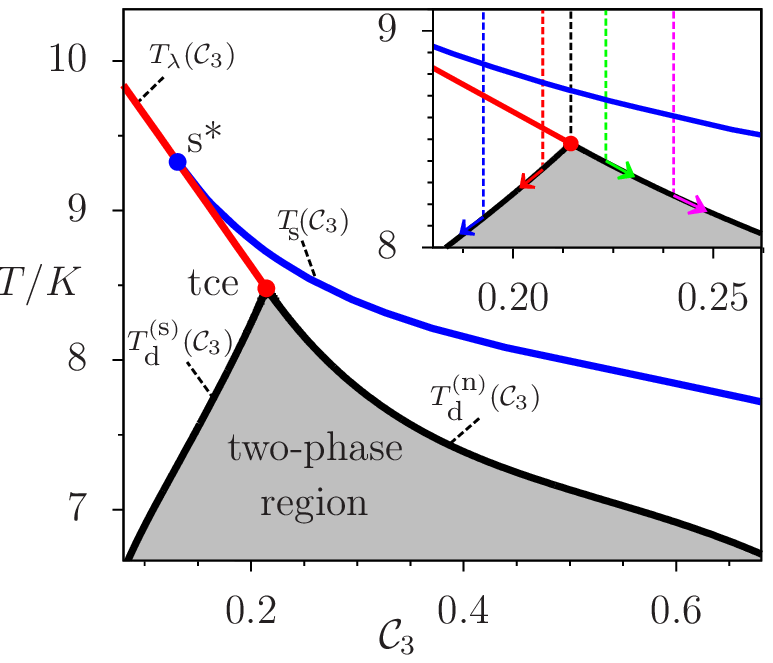}
\centering
\centering
\caption{\linespread{0.8}
\footnotesize{The bulk
liquid--liquid
phase transitions
at coexistence with the vapor phase
as in Fig.~\ref{different_topologies} (c)
plotted
in the $(T,\mathcal{C}_3)$ plane,
with $\mathcal{C}_3=(D-X)/2D$ as the
concentration
of $\Ht$
(the vapor phase is not shown here).
The blue line
$T_\text{s} (\mathcal{C}_3)$
represents
the
continuous
surface transition. Upon crossing this transition line a thin film near the wall
becomes superfluid although the bulk remains a normal fluid.
This line merges with 
the $\lambda$-line
(red line denoted as $T_\lambda (\mathcal{C}_3)$)
at the special point s*. The inset 
shows the
vertical
thermodynamic paths
(at liquid--vapor coexistence)
taken experimentally.
The numerical paths in our calculations are located in the vapor phase
parallel to the ones in the inset.
$T_\text{d}^{(\text{s})} (\mathcal{C}_3)$ $\left[T_\text{d}^{(\text{n})}(\mathcal{C}_3)\right]$
denotes the superfluid $[\text{normal fluid}]$ binodal of the two--phase region.
The
arrows
indicate how the vertical thermodynamic paths continue
after
encountering
the demixing curve.
The path shown by the black dotted line can follow both binodals.
}}
\label{bulk-modify}
\end{figure}

\begin{figure}
\includegraphics[width=85mm]{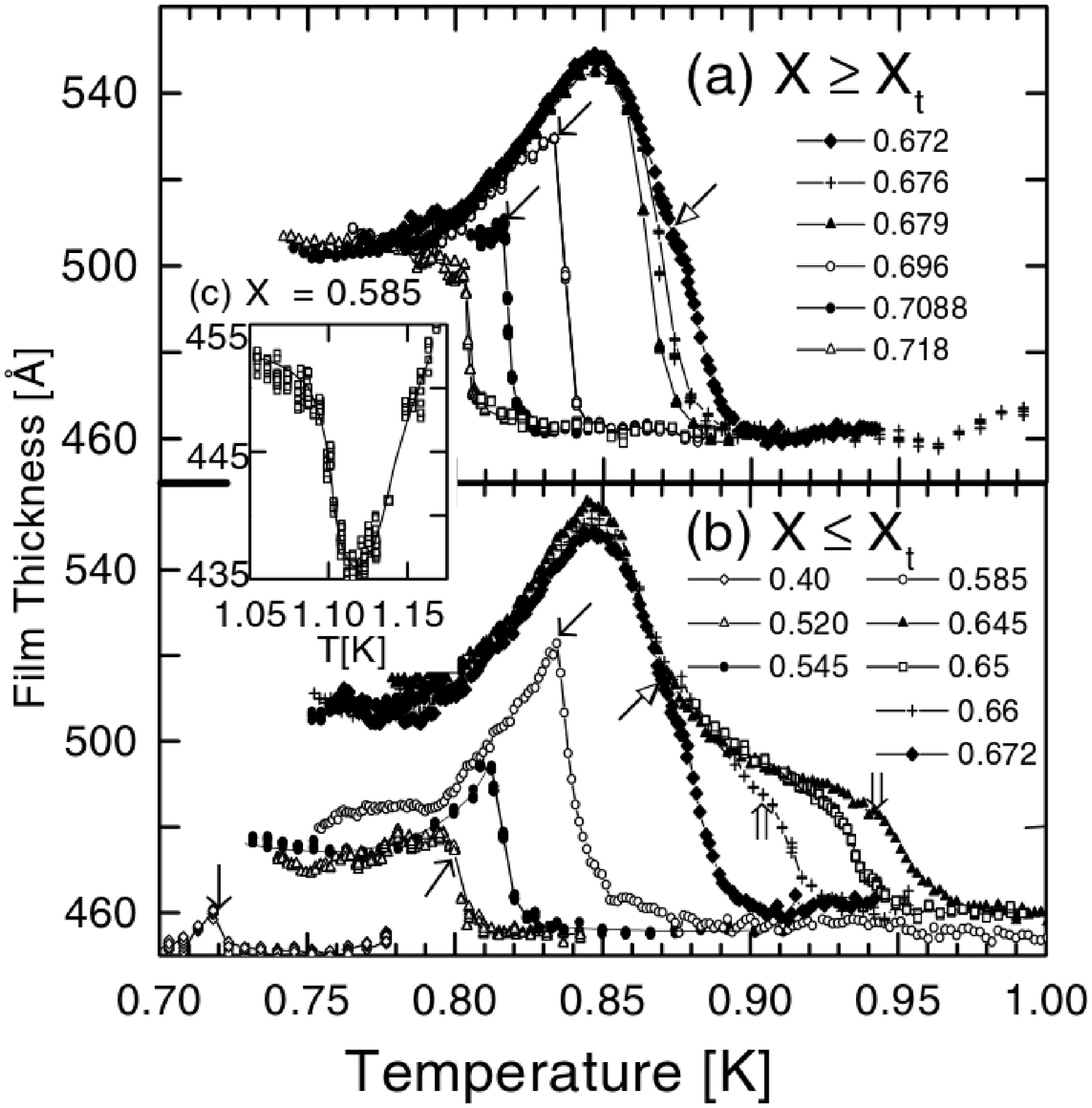}
\centering
\centering
\caption{\linespread{0.8}
\footnotesize{Thickness of 
 $\Ht$ - $\Hf$ wetting  films extracted from capacity measurements  (Fig. 4 in   Ref.~\cite{indirect}).
The
values of
$X$ refer to
various concentrations of $\Ht$.
The concentration of $\Ht$  at the tricritical point is $X_\text{t}=0.672$.
Panels (a) and (b) corresponds to $X\ge X_\text{t}$ and $X\le X_\text{t}$, respectively.
Thin arrows show the points, where the bulk liquid phase
separates.
The
large headed arrow indicates the tricritical point.
In (b)
the arrows with double lines
indicate the onset temperature of superfluidity.
For the thermodynamic path on the superfluid side (panel (b)), the growth
of the film thickness exhibits a characteristic shoulder 
between the tricritical
temperature
and the superfluid transition temperature
on the $\lambda$--line.
The growth of the film thickness
as a function
of temperature
is due to repulsive
TCFs
between the
solid wall
and the liquid-vapor interface, arising near the tricritical point.
For small $X$ the wetting film resembles
a film of pure $\Hf$, which corresponds to $(O,O)$ BCs
for the CCFs arising near
the temperature of the $\lambda$ transitions, which are
attractive~\cite{Chanhe4} (see the dip in panel (c)).
\REDCOLOR{Reprinted figure with permission from Ref.~\cite{indirect}.}
}}
\label{Indirect-Fig}
\end{figure}

\begin{figure}
\includegraphics[width=100mm]{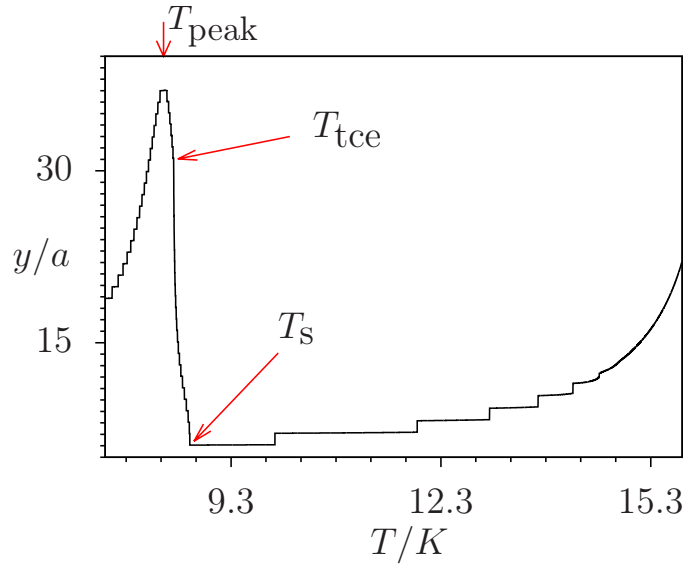}
\centering
\centering
\caption{\linespread{0.8}
\footnotesize{Numerical results
for the film thickness
corresponding to the thermodynamic path at fixed $\mathcal{C}_3=\mathcal{C}_3^{\text{tce}}$
and $\Delta \mu_+/K=-1.07\times 10^{-4}$ (i.e.,
slightly shifted
thermodynamic path shown by
the
vertical
black dashed line in
Fig.~\ref{bulk-modify}).
The arrows indicate the tricritical end point $T_\text{tce}$
and the onset temperature $T_\text{s}\simeq T_\text{s}(\mathcal{C}_3)$ for superfluidity
at the surface transition.
(The deviation of $T_\text{s}$ from
$T_\text{s}(\mathcal{C}_3)$
(see Fig.~\ref{bulk-modify})
is due to the offset from liquid--vapor coexistence.)
Below the tricritical temperature
the thermodynamic path follows $T_{\text{d}}^{(\text{s})}(\mathcal{C}_3)$
indicated in Fig.~\ref{bulk-modify} (infinitesimally on the superfluid side).
For further
discussions
see the main text.
The bulk parameters of the system are those belonging to
Fig.~\ref{different_topologies}(c) and Fig.~\ref{bulk-modify}.
$T_\text{peak}/K=8.3346$ is the position of the peak.
}}
\label{thickness_all_T}
\end{figure}

\begin{figure}
\includegraphics[width=100mm]{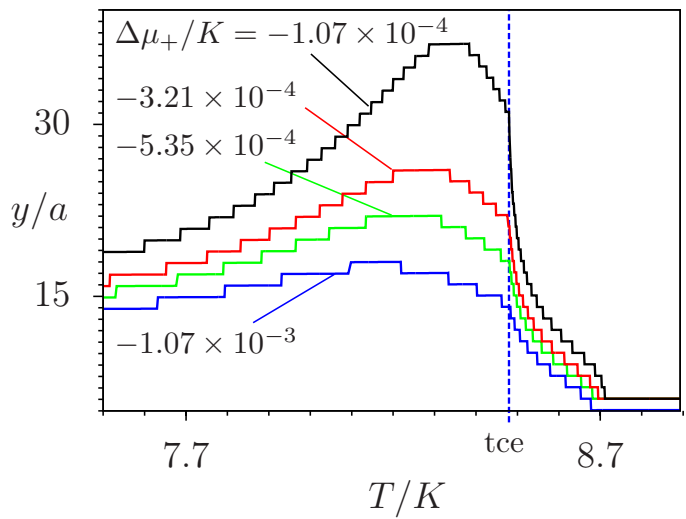}
\centering
\centering
\caption{\linespread{0.8}
\footnotesize{Film thickness versus temperature at $\mathcal{C}_3=\mathcal{C}_3^{\text{tce}}$ for
four
values of
$\Delta \mu_+/K$.
By increasing the offset value
$|\Delta\mu_+|$
the tricritical Casimir effect
and complete wetting become
less pronounced.
}}
\label{Xtce_different_offsets}
\end{figure}

\begin{figure}
\includegraphics[width=100mm]{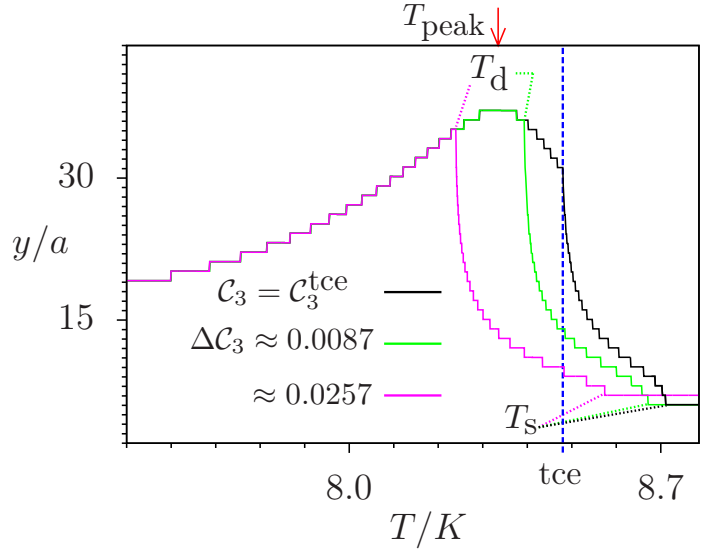}
\centering
\centering
\caption{\linespread{0.8}
\footnotesize{Film thickness $y/a$
as function of
temperature  $T$ for three
values of $\mathcal{C}_3\geq\mathcal{C}_3^{\text{tce}}$,
i.e.,
$\Delta\mathcal{C}_3=\mathcal{C}_3-\mathcal{C}_3^{\text{tce}}\geq 0$,
and at
$\Delta \mu_+/K=-1.07\times 10^{-4}$.
The sudden drop in the green and
in
the violet
curve occurs
at $T_\text{d}$ close to the
demixing temperature
$T_{\text{{d}}}^{(\text{n})}(\mathcal{C}_3)$ (see Fig.~\ref{bulk-modify}
with the same color code).
Below
$T_{\text{{d}}}^{(\text{n})}(\mathcal{C}_3)$
the
violet
and the green
curve
merge
with
the black curve
and follow the binodal denoted by
$T_{\text{{d}}}^{(\text{n})}(\mathcal{C}_3)$ in Fig.~\ref{bulk-modify}.
The black curve is similar to the one in Fig.~\ref{thickness_all_T}
except that below $T_\text{tce}$ it follows the normal branch of the binodal
(see Fig.~\ref{different_topologies}(c)).
The jumps are due to
first--order
layering transitions.
This figure corresponds to panel (a) in Fig.~\ref{Indirect-Fig}.
Note that $X$ in Fig.~\ref{Indirect-Fig} corresponds to $\mathcal{C}_3$
here
and $X_{\text{t}}$ corresponds
to $\mathcal{C}_3^{\text{tce}}$
here.
Due to the offset from liquid--vapor coexistence
the values of $T_\text{d}$ and $T_\text{s}$
differ slightly from $T_{\text{{d}}}^{(\text{n})}(\mathcal{C}_3)$ and $T_\text{s}(\mathcal{C}_3)$
as
shown
in Fig.~\ref{bulk-modify}.
The bulk parameters of the system are the same as in
Fig.~\ref{different_topologies}(c) and Fig.~\ref{bulk-modify}.
}}
\label{film_thickness_N}
\end{figure}

In order to
describe
wetting films of $\Ht$ -$\Hf$ mixtures, we
focus on systems exhibiting phase diagrams
with nonzero values of $J_s$ as in Fig.~\ref{different_topologies}(c) and
we
choose the surface fields
$(\tilde{f_+},\tilde{f_-})/K=(10.714,16.071)$.
All figures in this subsection
(i.e., Figs.~\ref{changing_offset_X3_constant}~-~\ref{film_thickness_N})
\RED{share the coupling constants $(C/K, J/K, J_s/K)=(1, 9.10714, 3.70107)$,
the surface fields $(\tilde{f_+},\tilde{f_-})/K=(10.714,16.071)$,
and the system size $L=60$.}
The growth of wetting films upon approaching the liquid-vapor coexistence
surface is
illustrated
in
Fig.~\ref{changing_offset_X3_constant}, where we have used  Eq.~(\ref{thicknessDefinition})
for defining
the film thickness.
For all temperatures considered,
upon approaching
liquid--vapor coexistence
the wetting films become thicker:
$y(\Delta_+\to0^-)\sim\kappa\text{ln}\frac{1}{|\Delta\mu_+|/K}$
with
a
significant temperature dependence of
the amplitude
$\kappa$.
This is different from the situation in Fig.~\ref{changing-offset-normal}
with $J_s=0$,
where only for
sufficiently high temperatures (i.e., $T>T_w$) complete wetting occurs.
This means that
in Fig.~\ref{changing_offset_X3_constant}
$T_w$ is below the considered temperature interval.
Interestingly,
in Fig.~\ref{changing_offset_X3_constant}
at the
reduced
temperature  $(T-T_{\text{tce}})/T_{\text{tce}}\approx -0.016$
the film thickness
exhibits
the most rapid increase upon approaching the liquid-vapor coexistence surface (see the red curve),
whereas for higher and lower temperatures the growth of the 
film thickness is
reduced, i.e., the amplitude $\kappa(T)$ introduced above has a maximum at $(T-T_{\text{tce}})/T_{\text{tce}}\approx -0.016$.
This is different from what one observes in Fig.~\ref{changing-offset-normal},
where the
thickness
of the wetting film is,
via $\kappa(T)$, a monotonically increasing function of $T$.
Note that
in Fig.~\ref{changing_offset_X3_constant}
for the curves with $T\geq T_\text{tce}$
the number density of $\Ht$ is fixed at $X_3=X_3^{\text{tce}}=0.20845$.
However, for $T< T_\text{tce}$ the system phase separates and the
number density of $\Ht$ changes. Accordingly,
in Fig.~\ref{changing_offset_X3_constant} for $(T-T_\text{tce})/T_\text{tce}=-0.016$
and $(T-T_\text{tce})/T_\text{tce}=-0.042$, the number density of
$\Ht$
on the superfluid branch of the binodal (Fig.~\ref{different_topologies}(c))
is $X_3=0.201$ and $X_3=0.188$, respectively.
The OP profiles for three temperatures
at $\Delta \mu_+/K=-1.07\times 10^{-4}$
are shown in
Fig.~\ref{profiles_at_three-T}.
Due to the large value
of $\tilde{f_-}$, the
number density
$X_{4,l}$
of $\Hf$ is enhanced near the wall
and hence
$X_l=X_{4,l}-X_{3,l}$
is large there.
If
the bulk liquid  is in the normal fluid phase but close to
either
the
$\lambda$-line
for $T>T_{\text{tce}}$, or to
the
normal branch of the binodal (Fig.~\ref{different_topologies}(c))
for $T<T_{\text{tce}}$, this
enhancement induces
symmetry breaking of the superfluid OP near the wall.
At  the liquid-vapor coexistence surface, this so-called
surface transition occurs at temperatures $T_{\text{s}}(X_3)$,
which depend on the
bulk
number density
$X_{3}$
of $\Ht$ atoms
or,
equivalently,
on the
bulk
concentration
$\mathcal{C}_3$
of $\Ht$ as $T_{\text{s}}(\mathcal{C}_3)$
(see Fig.~\ref{bulk-modify}).
With the bulk being in
the vapor phase, the
continuous
surface transition occurs within the wetting film 
for offsets  $\Delta \mu_{+}$
from the liquid-vapor coexistence surface smaller than a certain
temperature dependent
value,
which is marked 
in Fig.~\ref{changing_offset_X3_constant} by the
tick
along the
abscissa colored accordingly.
Upon crossing  the
continuous
surface transition one observes a nonzero profile $M_l$ in the wetting film
(see Figs.~\ref{profiles_at_three-T}(a) and (b)).
For
$T<T_{\text{tce}}$,
for which
the bulk liquid phase separates into
a superfluid and a normal fluid phase, 
the OP profiles within the  wetting films exhibit two plateaus,
one corresponding to the superfluid phase
(note the
left
plateau of $M_l$ in Fig.~\ref{profiles_at_three-T}(c))
and the other one
(on the right side)
corresponding to the normal fluid phase.
The minimum of
the profile $X_l$
occurs at the
emerging
liquid-vapor interface at around
(a) $l=31$, (b) $l=9$, and (c) $l=17$. This demonstrates
the effective attraction  of  $\Ht$ 
towards the
emerging
liquid-vapor
interface,
which suppresses the superfluid OP at the liquid--vapor interface. On the other hand
the preference of the wall for $\Hf$ enhances the superfluid OP there as if there would be
a surface field acting on the superfluid OP, which is , however, not the case.

The experimental data~\cite{indirect}, reproduced in Fig.~\ref{Indirect-Fig},
have been obtained at liquid-vapor coexistence along the paths of fixed  concentration
$\mathcal{C}_3$
of $\Ht$ 
as shown  in the inset of
Fig.~\ref{bulk-modify}
by the vertical dotted lines.
(Note that in Fig.~\ref{Indirect-Fig}
$X$
corresponds to the
concentration of $\Ht$, which here is denoted by
$\mathcal{C}_3=(D-X)/(2D)=X_3/(X_3+X_4)$.
We have ignored the subscript $l$ because
we are referring to the bulk values.)
The thermodynamic
paths of fixed $\Ht$ concentration followed in our calculations are parallel to the experimental ones 
but are located in the  vapor phase
close to
the
liquid-vapor coexistence surface
(like the brown surface in Fig.~\ref{3d_no_path}). 

The film thickness versus  temperature along
a
path
with an offset $\Delta \mu_+/K=-1.07\times 10^{-4}$
parallel to the
vertical
black dashed line in Fig.~\ref{bulk-modify}
is shown in Fig.~\ref{thickness_all_T}.
Within the considered temperature range the system is above the wetting temperature $T_{\text{w}}$ (not shown in the figure).
We
find
that  at fixed  $\mathcal{C}_3$ the  variation of the film thickness with temperature is nonmonotonic.
Upon increasing the
temperature, for $T>T_\text{s}$, the
film thickness increases.
A much steeper increase of the
film thickness, associated with a
break in
slope,
occurs
between $T_\text{s}$ and $T_\text{tce}$,
where the  TCFs emerge.
(Note that due to the offset from
liquid--vapor coexistence
the sharp drop
of $y/a$
occurs slightly
below
$T_\text{tce}$
(see Fig.~\ref{bulk-modify}).)

As discussed before,  due to the surface transition close to $T_\text{tce}$ 
 the superfluid OP    becomes nonzero near the wall.
This profile vanishes at the
emerging
liquid-vapor interface, where the  $\Ht$ atoms accumulate.
This
behavior corresponds 
to  the 
non-symmetric,
effective
$(+,O)$ BCs for the superfluid order parameter $M_l$ in the wetting film.
Therefore, the resulting  TCF acting on the liquid-vapor interface is repulsive and leads to
an increase of
the
film thickness.
The maximum film thickness
occurs at $T_{\text{peak}}/K\approx 8.3346$, which lies below $T_{\text{tce}}$ - in agreement
with the experimental results (see Fig.~\ref{Indirect-Fig})
($T_{\text{peak}}$ is defined as the mid point of the temperature range
enclosing
the maximum film thickness).
$T_{\text{s}}$ denotes the temperature of the surface transition. 
Figure~\ref{Xtce_different_offsets}
shows how the offset value $\Delta\mu_+$ affects the equilibrium film thickness $y$.
As expected,
upon increasing the offset value, the
film thickness decreases. Moreover $T_{\text{peak}}$ shifts towards lower temperatures.

Following the other thermodynamic paths
indicated
in Fig.~\ref{bulk-modify}
renders a
distinct
scenario.
Figure~\ref{film_thickness_N}
shows the
film thickness $y$ versus temperature $T$ for two values of
$\mathcal{C}_3>\mathcal{C}_3^{\text{tce}}$
(green curve and violet curve)
at $\Delta \mu_+/K=-1.07\times 10^{-4}$.
(As a reference, we plot also
the results for $\mathcal{C}_3=\mathcal{C}_3^{\text{tce}}$ (black curve)).
The maximum of
each of these two
curves
occurs at
a
temperature  close to  the corresponding bulk demixing temperature
denoted as
$T_{\text{d}}^{(\text{n})}(\mathcal{C}_3)$.
(This slight
deviation from $T_{\text{d}}(\mathcal{C}_3)$
is due to the offset
$\Delta\mu_+$
from the liquid-vapor coexistence surface.)
The green curve corresponding to
$\Delta \mathcal{C}_3=\mathcal{C}_3-\mathcal{C}_3^{\text{tce}}=0.0087$
joins the black one at
$T/K\simeq T_{\text{d}}^{(\text{n})}(\mathcal{C}_3)/K=8.3925$;
for lower temperatures both curves merge.
Since for the green curve
$T_{\text{d}}^{(\text{n})}(\mathcal{C}_3)>T_{\text{peak}}=8.3346$,
the maximum of this  curve is the same as the maximum of  the black curve.
However, for $\Delta \mathcal{C}_3=0.0257$ the violet curve joins the black curve at the corresponding
demixing temperature
$T_{\text{d}}^{(\text{n})}(\mathcal{C}_3)/K=8.2392$,
which is below the temperature
$T_\text{peak}$
of the peak. Therefore, the maximum of the  violet curve
differs
from the maximum of the black curve.
Figure~\ref{film_thickness_N}
corresponds to panel (a) in Fig.~\ref{Indirect-Fig}.
Note that $X_{\text{t}}$ in Fig.~\ref{Indirect-Fig} corresponds
to $\mathcal{C}_3^{\text{tce}}$
in the present notation.

\begin{figure}
\includegraphics[width=95mm]{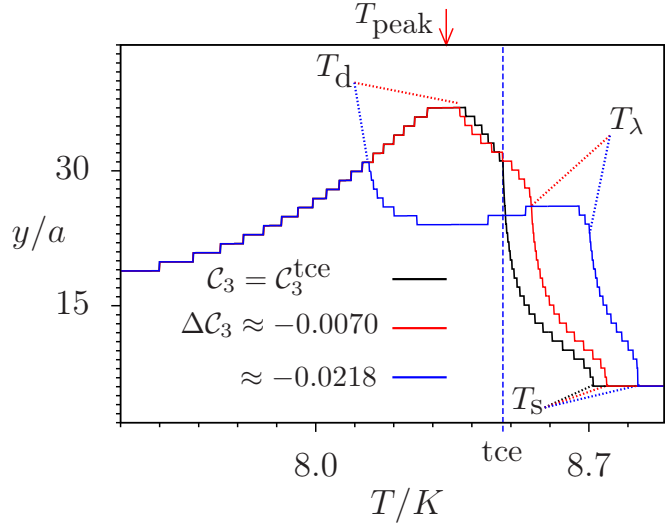}
\centering
\centering
\caption{\linespread{0.8}
\footnotesize{Film thickness $y/a$
as function of
temperature  $T$ for three
values of $\mathcal{C}_3\leq\mathcal{C}_3^{\text{tce}}$
with
$\Delta\mathcal{C}_3=\mathcal{C}_3-\mathcal{C}_3^{\text{tce}}\leq0$
and at
$\Delta \mu_+/K=-1.07\times 10^{-4}$.
The black curve is the same as the one in Fig.~\ref{thickness_all_T}.
The sudden drop in the blue and
in the red curve occurs
at $T_\lambda$ close to
the temperature of the $\lambda$-transition $T_{\lambda}(\mathcal{C}_3)$
(see Fig.~\ref{bulk-modify}).
The red and the blue curve merge with the
black curve at $T_{\text{{d}}}^{(\text{s})}(\mathcal{C}_3)$
and follow the binodal denoted by
$T_{\text{{d}}}^{(\text{s})}(\mathcal{C}_3)$ in Fig.~\ref{bulk-modify}.
This figure corresponds to panel (b) in Fig.~\ref{Indirect-Fig}.
Note that $X$ in Fig.~\ref{Indirect-Fig} corresponds to $\mathcal{C}_3$
here
and $X_{\text{t}}$ corresponds
to $\mathcal{C}_3^{\text{tce}}$
here.
Due to the offset from liquid--vapor coexistence
the value of $T_\text{d}$, $T_\text{s}$, and $T_\lambda$
differ slightly from $T_{\text{{d}}}^{(\text{n})}(\mathcal{C}_3)$, $T_\text{s}(\mathcal{C}_3)$,
and $T_\lambda(\mathcal{C}_3)$
as
introduced
in Fig.~\ref{bulk-modify}.
The bulk parameters of the system are the same as in
Fig.~\ref{different_topologies}(c) and Fig.~\ref{bulk-modify}.
}}
\label{film_thickness_S}
\end{figure}

Figure~\ref{film_thickness_S}
shows the
film thickness
as function of
temperature for two  values of $\mathcal{C}_3<\mathcal{C}_3^{\text{tce}}$
(red curve and blue curve; compare the inset in Fig.~\ref{bulk-modify} with the same color code)
and  for $\mathcal{C}_3^{\text{tce}}$  (black curve)
at
$\Delta \mu_+/K=-1.07\times 10^{-4}$.
The blue curve and the red curve merge with the black one
at $T_\text{d}$ close to the
demixing temperature denoted as $T_{\text{d}}^{(\text{s})}(\mathcal{C}_3)$ in Fig.~\ref{bulk-modify}
(there is a slight deviation due to the offset from the liquid--vapor coexistence).
Whereas for  $\mathcal{C}_3 \ge \mathcal{C}_3^{\text{tce}}$ the sudden  drop of the
film thickness occurs  near
$T_{\text{d}}^{(\text{n})}(\mathcal{C}_3)$
(note that for
$T_{\text{d}}^{(\text{s},\text{n})}(\mathcal{C}_3^{\text{tce}})=T_\text{tce}$),
for $\mathcal{C}_3<\mathcal{C}_3^{\text{tce}}$ it takes place
close to the bulk $\lambda$-transition temperature
$T_{\lambda}(\mathcal{C}_3)\geqslant T_\text{tce}$.
(Again, there is a slight
deviation
due to the offset from the liquid-vapor coexistence surface.)
This sudden  drop is associated with
a break in
slope in the curves $y(T)$ and   leads to the formation of  characteristic shoulders.
This
agrees with the experimental observations (see panel (b) in Fig.~\ref{Indirect-Fig}).
Note that because $T_{\lambda}(\mathcal{C}_3)$ is a decreasing function of $\mathcal{C}_3$,
for lower concentrations of $\Ht$, the break in slope occurs at higher temperatures.
For the red curve in Fig.~\ref{film_thickness_S}, this shoulder is due to the emerging of the CCFs close to the $\lambda$-line.
For even lower values of $\mathcal{C}_3$ the films encounter only the CCFs due to
the $\lambda$--transition
and the TCFs due to the tricritical point do not influence them (see the blue curve).
In Fig.~\ref{film_thickness_S} all curves attain their lowest value at the surface transition
temperature
$T_{\text{s}}(\mathcal{C}_3)>T_\lambda(\mathcal{C}_3)$.

For a vertical path at $\mathcal{C}_3<\mathcal{C}_3^{\text{s*}}$
(see Fig.~\ref{bulk-modify}),
the film thickness does not
exhibit
an increase near the
${\lambda}$--transition.
In fact, for  $\mathcal{C}_3<\mathcal{C}_3^{\text{s*}}$ the BCs
for
the  superfluid OP
at the
interface
of the  wetting film are the symmetric (O,O) BCs
(i.e., $M=0$
at the wall  and at the
emerging
liquid-vapor interface).
Therefore, in this regime
one expects the
occurrence
of  an
\textit{attractive}
CCF; however,
this cannot be captured within
the present mean field approximation
because for Dirichlet--Dirichlet BCs the resulting CCF is solely
due to fluctuations beyond
mean field theory~\cite{prl66,1886}.
Although both black curves in Fig.~\ref{film_thickness_N}~and~\ref{film_thickness_S}
correspond to $\mathcal{C}_3=\mathcal{C}_3^{\text{tce}}$, they differ
slightly
due to the infinitesimal difference of the thermodynamic
paths for $T<T_{\text{tce}}$.
In Fig.~\ref{film_thickness_N},
for $T<T_{\text{tce}}$ 
the thermodynamic paths follow the demixing line
$T_{\text{d}}^{(\text{n})}(\mathcal{C}_3)$ infinitesimally on the normal fluid side, whereas in
Fig.~\ref{film_thickness_S} for $T<T_{\text{tce}}$ the thermodynamic paths
run along the superfluid binodal
$T_{\text{d}}^{(\text{s})}(\mathcal{C}_3)$.



\subsection{Tricritical Casimir Forces}
\label{subsec:tcf}
A fluid film exerts an effective force on its confining walls. For
two parallel,
planar walls a distance $L$ apart
this fluid mediated force $f_s$
is given by~\cite{Evans:1990}
\begin{align}
\label{CasimirDefinitionSlab}
 f_\text{s}=-\left(\frac{\partial {\cal F}^{\text{ex}}}{\partial L}\right)_{T,\,\mu} 
 =-\left(\frac{\partial({\cal F} -V f_\text{b})}{\partial L}\right)_{T,\,\mu},
\end{align}
where $f_\text{b}$ is the
grand canonical
bulk free energy density of a
one--component
fluid at temperature $T$ and chemical potential $\mu$.
${\cal F}$ is the free energy of the
film of volume
$V={\cal A}L$ where ${\cal A}$ is the macroscopically 
large surface area of one wall. 
Since $\mathcal{F}-Vf_b$ is proportional to $\mathcal{A}$,
$f_\text{s}/{\cal A}$ is the pressure
in excess
over its bulk value.
Upon approaching the bulk critical point of the
confined
fluid,
$f_\text{s}$ acquires
a
\emph{universal}
long-ranged contribution $f_C$, known as the critical Casimir force~\cite{Krech:1997,danchev,Gambassi:2009}.

Extending this concept to binary liquid mixtures, here
we focus on that
contribution to  $f_\text{s}/{\cal A}$
which arises near a
tricritical point
of $\Ht$ - $\Hf$ mixtures. We call this contribution
tricritical Casimir force $f_\text{tcr}$ (TCF)
and express it in units of $k_{\text{B}}T_{\text{tc}}$, where
$T_{\text{tc}}$ is the temperature of a
tricritical point
on the line TC in Fig.~\ref{3d_no_path}.

As discussed in the
Introduction, concerning wetting
by
a critical fluid,
the critical fluctuations
of the OP
are confined by
the solid
substrate
surface on one side and by the
emerging
liquid-vapor interface on the other side.
Accordingly, the TCF
is
the derivative 
of  the corresponding  excess
free energy with respect  to  the  film thickness $y$ at constant temperature and chemical potentials.
In contrast to  the slab geometry
with two fixed walls as discussed above (see  Eq.~(\ref{CasimirDefinitionSlab})),
varying the equilibrium wetting
film thickness
requires to change
the thermodynamic state of
the
fluid. Moreover, in the present microscopic approach the film thickness
is not
an input
parameter of a model; hence, the excess free energy is not an explicit function of $y$.
(Note
that  $y$ is uniquely defined
in terms of
the equilibrium
density profile $D_l(T,\mu_+,\mu_-)$ via Eq.~(\ref{thicknessDefinition}).)
In order to calculate the TCF,
we consider a system at fixed
$T, \mu_+$, and $\mu_-$, for
which the film thickness is fixed 
to a specific  value $\ell$ by an
externally imposed
constraint.
For the total
free energy $F_\text{cstr}$ of such
a constraint
system, one has for large $L$~\cite{generalphasediagram,1886,dietrich-napiorkowski-1991}
\begin{equation}
\label{totalFreeContributions_constraint}
 F_\text{cstr}(T,\mu_+,\mu_-,\ell)/A = f_\text{m}\ell+f_\text{b}(L-\ell)+\sigma_\text{w,l}+\sigma_\text{l,v}+f_\text{ex}(\ell),
\end{equation}
where $\sigma_\text{w,l}$ and
$\sigma_\text{l,v}$
are the wall-liquid and vapor-liquid
surface tensions, respectively, $f_\text{m}$ is the free energy density of the metastable liquid, and 
$A:=\mathcal{N}a^2$ is
the
cross section area of a layer.
Since at liquid--vapor coexistence $f_\text{b}=f_\text{vapor}=f_\text{liquid}<f_\text{m}$
one has
$(f_\text{m}-f_\text{b})l>0$.
The $\ell$-dependent excess free energy $f_\text{ex}(\ell)$
is the sum of
two contributions: the free energy density
(per area $A$)
$f_0(\ell)$  due to the
effective interaction of the emerging
liquid--vapor interface with the substrate wall
and  the
singular
contribution
$f_\text{sing}(\ell)$
due to the
critical
finite-size effects
within the wetting film of thickness $l$.
For
short--ranged
surface fields, the effective potential between
the wall and  the
emerging
liquid-vapor interface is an exponentially decaying function
of the film thickness $\ell$.
To leading order
one has~\cite{binder-landau-mueller-2002}
\begin{equation}
 \label{background}
 f_0(\ell)\approx\alpha \frac{T-T_w}{T_w} \exp(-p\ell),
\end{equation}
where $T_w$ is the wetting
transition
temperature
and $\alpha>0$
is
an amplitude such
that in accordance with complete wetting $f_0(l,T>T_w)>0$.
The decay length $1/p$ is the bulk correlation length of the liquid at $T_w$
and at liquid--vapor coexistence.
With the knowledge of $f_\text{ex}(l)$
and $f_0(l)$ one can determine the
TCF as
the
negative
derivative 
of $f_\text{ex}(\ell)-f_0(\ell)$
with respect to $\ell$.
Since $y(T,\mu_+,\mu_-)$ is the equilibrium film thickness,
the total free energy $F_\text{cstr}$
has a global minimum at $y$,
so that
$\frac{\partial F_\text{cstr}}{\partial \ell}|_{\ell=y}=0$.
Thus
taking
the derivative of
both sides of Eq.~(\ref{totalFreeContributions_constraint}) with respect to $\ell$ at $\ell=y$ yields
\begin{equation}
\label{derFconstraint}
 0 \approx f_\text{m}-f_\text{b}+\frac{\partial f_0}{\partial \ell}|_{\ell=y}+\frac{\partial f_{\text{sing}}}{\partial \ell}|_{\ell=y}.
\end{equation}
With Eq.~(\ref{background}) this implies for the TCF
\begin{equation}
\label{CasimirFormula}
 f_\text{TCF}(y)=-\frac{\partial f_{\text{sing}}}{\partial \ell}|_{\ell=y} \approx f_\text{m}-f_\text{b}-\alpha p\frac{T-T_w}{T_w} e^{-py} .
\end{equation}
The parameters  $\alpha$, $T_w$, and $p$
can be determined by studying   the growth of the equilibrium film thickness
as a function of the chemical potential sufficiently far above the critical
demixing
region, where  $f_\text{TCF}(y)$
is negligible.
Using Eq.~(\ref{CasimirFormula}) and calculating $f_\text{m}$
and $f_\text{b}$  within the present model,
we have found that for the surface fields
$(\tilde{f}_+,\tilde{f}_-)/K=(10.714,16.071)$
and the coupling constants $(C/K, J/K, J_s/K)=(1, 9.107, 3.701)$,
one has
$T_w/K\simeq3.704$, whereas $\alpha\simeq1.146$, and $p\simeq 1.997$.
We have checked that
the value of the bulk correlation length $1/p$
agrees with the one following
from the decay of the OP profiles.


In the slab geometry considered in Refs.~\cite{maciole-dietrich-2006,maciolek-gambassi-dietrich},
the
total number
density of
the
$\Ht$ -$\Hf$ mixtures  is fixed and the properties of the system  near the bulk tricritical
point  can be expressed in terms of
the experimentally accessible
thermodynamic fields $T-T_\text{tc}$
and $\mu_- -\mu^\text{tc}_-$, where $\mu^\text{tc}_-$ is the value of $\mu_-$ at the tricritical point. (The thermodynamic field conjugate to the superfluid OP is experimentally
not accessible and
is omitted here.)
As discussed in detail in Refs.~\cite{maciole-dietrich-2006,maciolek-gambassi-dietrich,riedel},
the proper dimensionless scaling fields are $t\equiv (T-T_\text{tc})/(T_\text{tc})$  and $g\equiv (\mu_- -\mu^\text{tc}_-)/(k_BT_\text{tc})+a't$,
where $a'$ is the slope of
the line tangential
to the phase boundary
curve
at $T_{\text{tc}}$
within
the blue surface in Fig.~\ref{3d_no_path}
(i.e., parallel to the intersection of
the blue surface and A$_4$ at tc
which is the full blue horizontal line through tc).
For such a choice of the scaling fields, for $t\to 0$ with $g=0$ the tricritical point 
is approached tangentially to the phase boundary.
According to finite-size scaling~\cite{FSS2}
the CCF for the slab
of
width $L$
is  governed by a universal scaling function 
defined as 
$\tilde{\vartheta}_{+,o} \simeq L^3f_\text{TCF}/(k_\text{B}T_{\text{tc}})$,
where the subscript 
$\{+,o\}$ denotes  the surface universality classes
of the confining surfaces (the symbol ``$\simeq$''
indicates
asymptotic equality).
The scaling function
$\tilde\vartheta_{+,o}$
depends on the
two scaling fields $c_1tL^{1/\nu}$
and $c_2gL^{\Delta/\nu}$, where $c_1$ and $c_2$ are nonuniversal metric factors and $\nu=1$ and $\Delta=2$ are tricritical  exponents 
for  the $XY$ model in $d=3$~\cite{sarbach}.
In order to facilitate a
comparison with experimental data,
the
results for the TCF obtained in
Refs.~\cite{maciole-dietrich-2006,maciolek-gambassi-dietrich}
have been presented in terms of $\tilde\vartheta_{+,o}$ as a function of only
the
single scaling variable $c_1tL^{1/\nu}$, with
$c_1 = \xi^{+}_0/a$; $\xi^{+}_0$
(in units of $a$) is the amplitude of the superfluid OP 
correlation length $\xi = \xi^{+}_0t^{-\nu}$ above $T_{\text{tc}}$.
In
Refs.~\cite{maciole-dietrich-2006,maciolek-gambassi-dietrich},
for thermodynamic paths of constant concentration,
the influence of the
variation of the second scaling
variable $g$ upon changing temperature
has been neglected.

In the present case of TCF emerging  in wetting films
of thickness $y$,
the TCF per area
is given by the universal scaling function $\mathcal{\vartheta}_{+,o}$ as
\begin{equation}
\label{scaling-casimir}
f_\text{TCF}/(k_\text{B} T_\text{tc})\simeq y^{-d} \mathcal{\vartheta}_{+,o}(c_1yt^{\nu})\text{,}
\end{equation}
where we have
again
neglected the dependence of $\mathcal{\vartheta}_{+,o}$ on the scaling variable
$c_2gy^{\Delta/\nu}$
as well as on the
third
scaling variable associated with $\mu_+-\mu_+^{\text{tc}}$
which is
conjugate
to the total number density of the $\Ht$ - $\Hf$ mixture. 
In order to retrieve,
however,
the full information stored in the scaling function,
in principle one has to plot the scaling function as a function of a single scaling variable,
while keeping
all
the other scaling variables fixed.
In practice this is difficult to realize.
Along the thermodynamic paths
taken experimentally
in
Ref.~\cite{indirect},
none of the scaling variables were fixed.
Instead
the scaling functions have been plotted versus the single scaling variable
$td$, where
in Ref.~\cite{indirect}
$d$
denotes
the film thickness.
We follow
this experimentally inspired approach
and
plot $y^3f_\text{TCF}/(k_\text{B} T_\text{tce})$  as a function of
$yt$, ignoring the nonuniversal metric factor $c_1$.
Since
the surfaces fields we have chosen for our calculation of the TCF are strong, we  neglect the 
dependence of the scaling function on the corresponding scaling
variables,
assuming that  for
$(\tilde{f_+},\tilde{f_-})/K=(10.714,16.071)$
the system is close to the fixed point $(+)$ BCs.

\begin{figure}
\includegraphics[width=120mm]{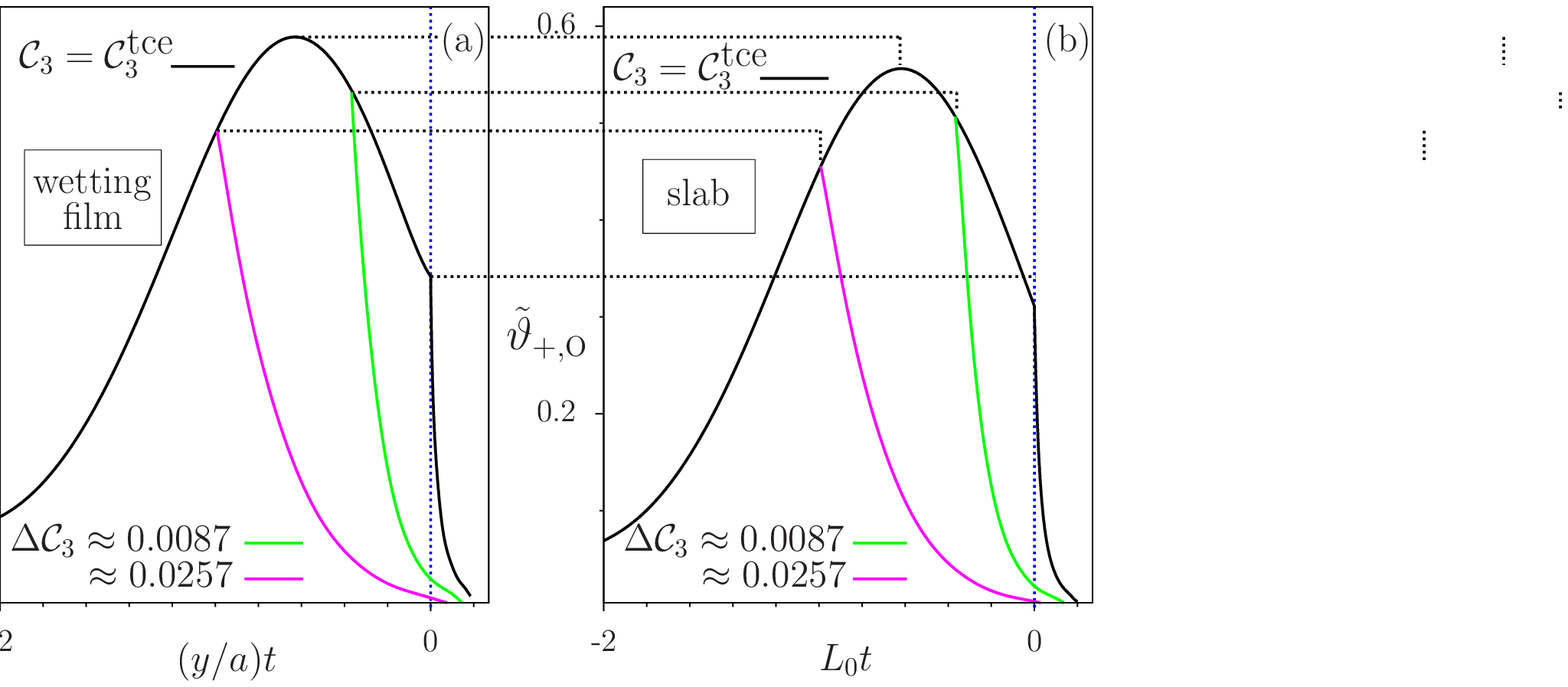}
\centering
\centering
\caption{\linespread{0.8}
\footnotesize{Scaling functions of
the
TCF
calculated from the data
in Fig.~\ref{film_thickness_N} within (a) the
wetting
film geometry
and (b) the slab approximation.
Concerning the definition of
the
slab thickness $L_0$ see the main text.
The magenta
curve
and the green
curve
merge with the black curve at their corresponding demixing point
indicated
by $T_\text{d}$ in Fig.~\ref{film_thickness_N},
using the same color code.
The corresponding curves in the two panels agree qualitatively
but differ in detail, e.g., in height
(see the horizontal lines).
In panel (a) the thermodynamic states
are off the liquid-vapor coexistence surface, whereas in panel (b) the thermodynamic
states
lie on
the liquid-vapor coexistence surface.
The reduced temperature is
$t=(T-T_\text{tce})/T_\text{tce}$, where $T_\text{tce}$ is the temperature
of the tricritical end point.
Due
to the smoothing procedure and within the
\tcr{presently available numerical accuracy,
the small difference between}
the positions of the
maxima in (a) and (b)
\tcr{cannot be resolved reliably.}
}}
\label{SLAB_SCALING_normal}
\end{figure}

\begin{figure}
\includegraphics[width=120mm]{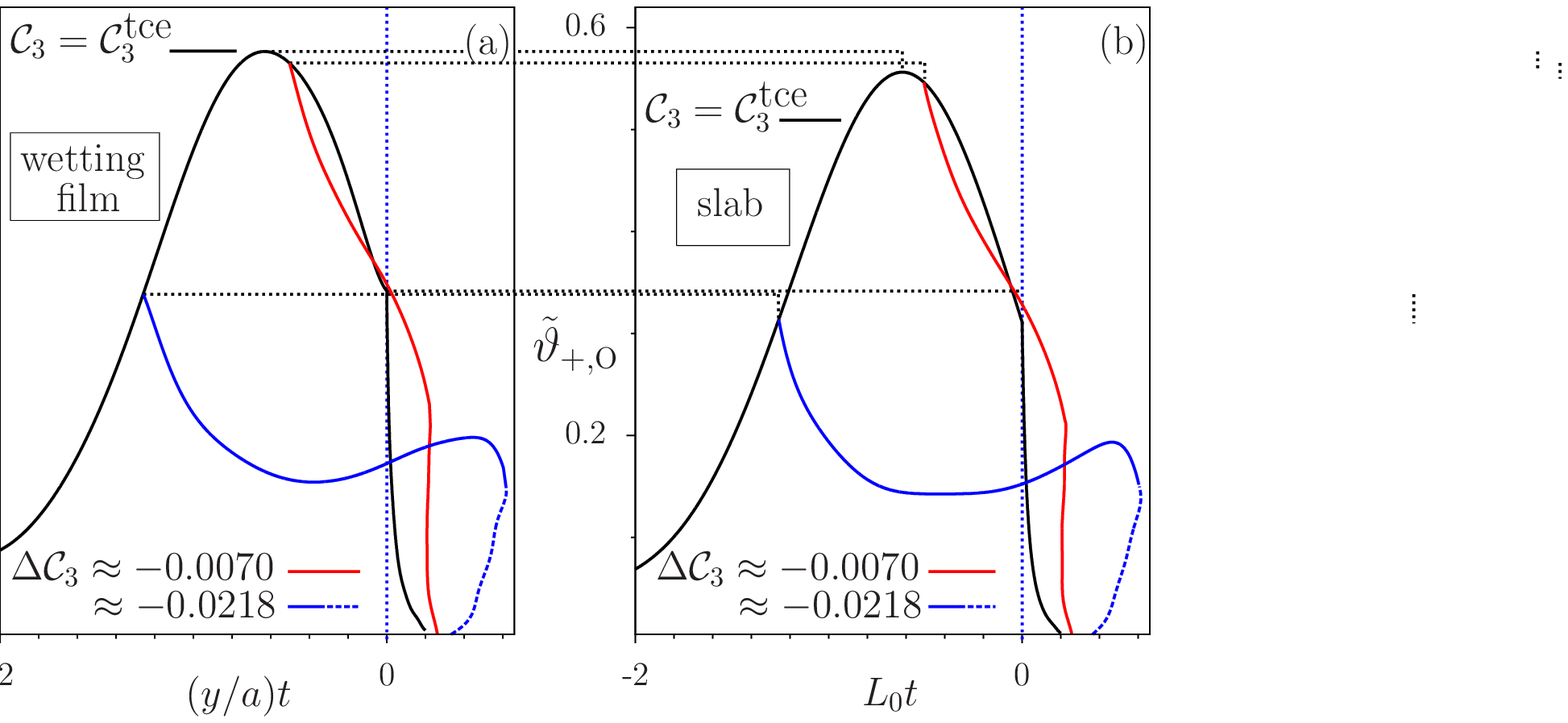}
\centering
\centering
\caption{\linespread{0.8}
\footnotesize{Scaling functions of
the
TCF
calculated from the data
in Fig.~\ref{film_thickness_S} within (a) the
wetting
film geometry
and (b) the slab approximation.
Concerning the definition of
the
slab thickness $L_0$ see the main text.
The blue curve and the red curve
merge with the black curve at their corresponding demixing point,
indicated
by $T_\text{d}$ in Fig.~\ref{film_thickness_N}.
The corresponding curves in the two panels agree qualitatively
but differ in detail, e.g., in height
(see the horizontal lines).
The dashed blue curve shows the region, where the blue curve is
multivalued
and scaling does not hold anymore. The same
holds
also for the
right parts of the red curves, because the drops of the curves exhibit a slightly
positive slope.
In panel (a) the thermodynamic states
are off the liquid-vapor coexistence surface, whereas in panel (b) the thermodynamic
states
lie on
the liquid-vapor coexistence surface.
The reduced temperature is
$t=(T-T_\text{tce})/T_\text{tce}$, where $T_\text{tce}$ is the temperature
of the tricritical end point.
Due
to the smoothing procedure and within the
\tcr{presently available numerical accuracy,
the small difference between}
the positions of the
maxima in (a) and (b)
\tcr{cannot be resolved reliably.}
}}
\label{SLAB_SCALING_super}
\end{figure}

\begin{figure}
\includegraphics[width=120mm]{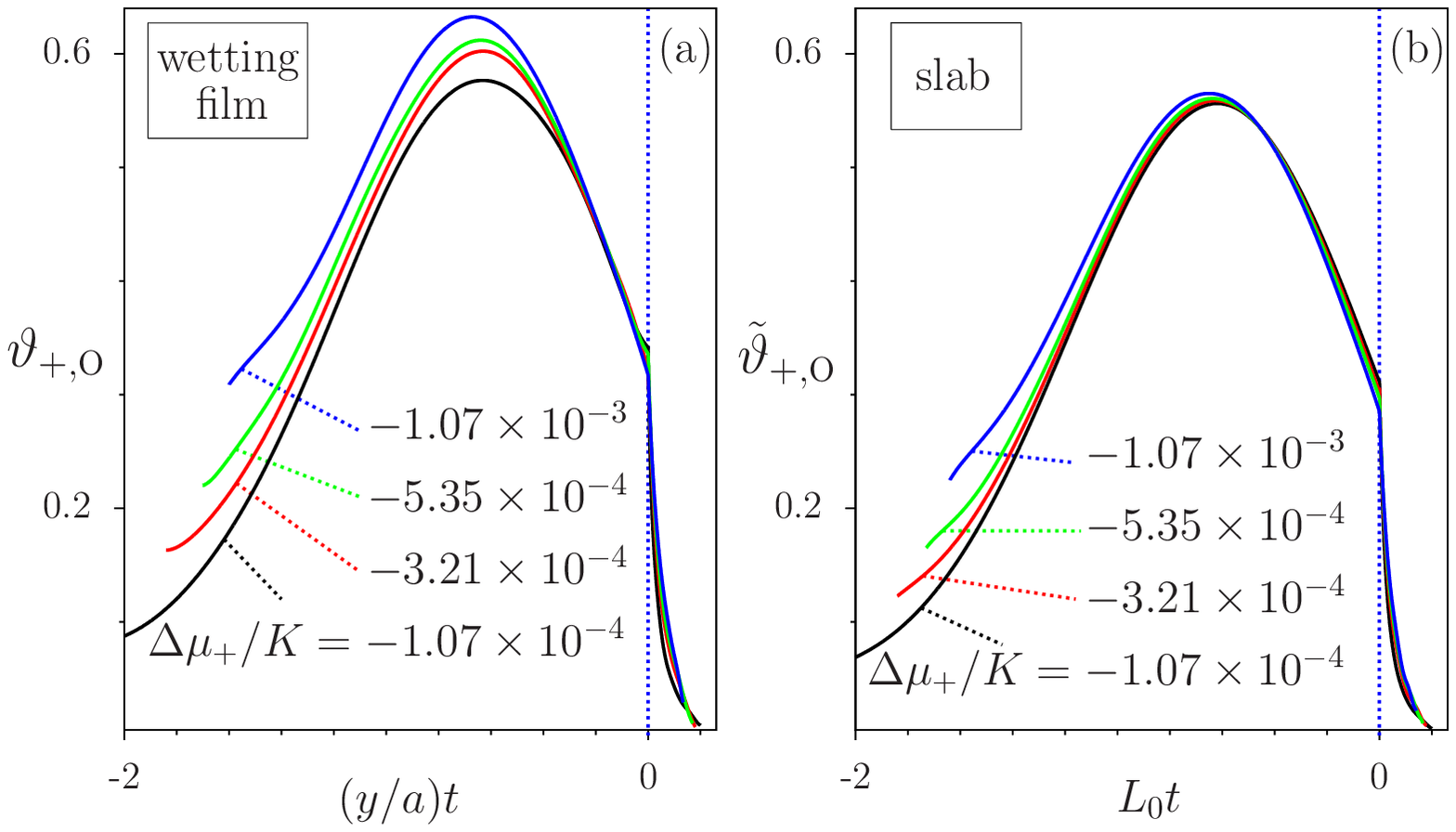}
\centering
\centering
\caption{\linespread{0.8}
\footnotesize{
Scaling functions of
the
TCF
calculated from the data
in Fig.~\ref{Xtce_different_offsets} within (a) the
wetting
film geometry
and (b) the slab approximation.
Concerning the definition of
the
slab thickness $L_0$ see the main text.
The maxima of the scaling functions in panel (a)
differ
from
each other,
whereas
the ones in panel (b) are almost
equal.
The reduced temperature is
$t=(T-T_\text{tce})/T_\text{tce}$, where $T_\text{tce}$ is the temperature
of the tricritical end point.
}}
\label{SLAB_SCALING_different_offsets}
\end{figure}

Figures~\ref{SLAB_SCALING_normal}(a)~and~\ref{SLAB_SCALING_super}(a)
show
the scaling functions calculated from the data
in Figs.~\ref{film_thickness_N}~and \ref{film_thickness_S}, respectively.
In order to
eliminate
the nonuniversal features arising from  the jumps
in the wetting films  due to the layering transitions, these curves have been smoothed. 
Figure~\ref{SLAB_SCALING_different_offsets}(a)
shows the scaling functions for
various values of $\Delta \mu_+$
corresponding to the various curves in
Fig.~\ref{Xtce_different_offsets}.
The vertical blue dotted line
in
Figs.~\ref{SLAB_SCALING_normal}~-~\ref{SLAB_SCALING_different_offsets}
represents the tricritical
end point
($t=0$).
Away from the tricritical temperature the scaling functions decay to zero. 
This decay  is faster for temperatures higher than  the tricritical temperature,
i.e., for $t>0$. For $t>0$ the dashed section
of the blue curve in Fig.~\ref{SLAB_SCALING_super}
shows
that
part, which
is multivalued.
This indicates that in this range of the scaling variable
the scaling hypothesis
is not applicable. The same holds also
for the
red curve in this figure, where the sudden drop
exhibits a slightly positive
slope.

In order to compare our
wetting
results for the TCF with those obtained in the slab geometry
as studied in Refs.~\cite{maciole-dietrich-2006,Maciolek-Gambassi-Dietrich-2007},
we employ  a suitable slab approximation
for our wetting data. To this end we consider
a slab
of
width $L_0$ equal to
the equilibrium position of the
emerging
liquid-vapor interface of the wetting film 
$L_0(T,\mu_+,\mu_-)=\lfloor y(T,\mu_+,\mu_-) \rfloor$,
at a certain
value of the offset $\Delta\mu_+$.
Since
within the present lattice model
the system size $L_0$ must
be
an integer,
the above assignment for $L_0$ involves the floor function $\lfloor\quad\rfloor$.
($\lfloor x\rfloor$
gives the largest integer number smaller than $x$.)

Within the slab approximation, the
emerging
liquid-vapor interface is replaced by
a wall (denoted by "2")
with the short-ranged surface fields
$\tilde{f}_{+,2}$ and $\tilde{f}_{-,2}$.
These surface fields are chosen such that the OP profiles calculated
for the slab
\textit{at} liquid--vapor coexistence (i.e., $\Delta\mu_+=0$) resembles the ones within the wetting
film geometry calculated for the semi--infinite system
with an
offset $\Delta\mu_+<0$.
In order to obtain a perfect match,
one
would have to allow these surface fields to vary along
the
thermodynamic paths
taken.
Insisting, however,
on fixed values of $(\tilde{f}_{+,2},\tilde{f}_{-,2})$, we have found that for
$(\tilde{f}_{+,2},\tilde{f}_{-,2})/K=(1.607,0.214)$
the profiles in the slab geometry agree
rather
well with their
counterparts in the
wetting
film
geometry.
For
$(\tilde{f}_{+,2},\tilde{f}_{-,2})/K=(1.607,0.214)$
the number density $X_{4,l}$
of $\Hf$ at the right boundary
is not
high
enough for the
spontaneous
symmetry breaking
of the superfluid OP to occur
there. On the contrary,
for $(\tilde{f}_{+},\tilde{f}_{-})/K=(10.714,16.071)$
at the left boundary
$M_l$ is nonzero.
Accordingly,
the two sets of surface fields
induce
$(+,O)$ and thus
non--symmetric
BCs on the superfluid OP within the
slab, giving rise to repulsive TCFs.
For such a slab,
by using Eq.~(\ref{CasimirDefinitionSlab})
we  calculate the TCF for
that bulk thermodynamic state which is associated with the wetting film, but taken
{\it at} bulk
liquid--vapor
coexistence
(i.e., $\Delta\mu_+\,=\,0$).
In this way we can mimic the
actual
experimental
wetting
situation
and stay consistent with the calculations for the slab geometry
as carried out
in Refs.~\cite{maciole-dietrich-2006,Maciolek-Gambassi-Dietrich-2007}.
Within lattice models,  the smallest change in the system size
amounts to
one layer ($\text{min}(\Delta L_0)=1$). Therefore, on the lattice the derivative
in Eq.~(\ref{CasimirDefinitionSlab}) has to be
approximated
by
the
finite difference
\begin{equation}
 \label{easySlab}
 f_\text{TCF}=-\frac{\Delta f^{\text{ex}}(L_0)}{\Delta L_0}=-(f^{\text{ex}}(L_0+1)-f^{\text{ex}}(L_0))
\end{equation}
where $f^{\text{ex}}={\cal F}^{ex}/A$. In order to determine  $f^{\text{ex}}(L_0)$,
we write
the total free energy $\phi$ of the slab within thickness $L_0$
as
\begin{equation}
 \label{contributionsSlab}
 \phi(L_0,T,\mu_+,\mu_-)/A=f_\text{b} L_0+\sigma_\text{s,l}^{(1)}+\sigma_\text{s,l}^{(2)}+f^\text{ex}(L_0),
\end{equation}
where
$\sigma_\text{s,l}^{(1)}$ and $\sigma_\text{s,l}^{(2)}$ are the surface tensions
between
the liquid and surface $(1)$ and surface $(2)$, respectively. The
surface tensions are functions of
$T,\mu_+,$ and $\mu_-$
only and
do not depend on the system size $L_0$.
Using Eq.~(\ref{contributionsSlab}),
Eq.~(\ref{easySlab}) can be expressed as
\begin{equation}
 \label{FinalSlab}
 f_\text{TCF}=(\phi(L_0)-\phi(L_0+1))/A+f_\text{b}.
\end{equation}
%

Figures~\ref{SLAB_SCALING_normal}(b),~\ref{SLAB_SCALING_super}(b),~and~\ref{SLAB_SCALING_different_offsets}(b)
show
the scaling functions $\tilde{\vartheta}_{+,o}$ within the slab
approximation,
corresponding to the cases in panel (a) of each figure.
Also here
curves have been  smoothed
out in order to
eliminate
the
discontinuities  due to the layering
transitions.
The approximation of the derivative in Eq.~(\ref{easySlab})
by a finite difference and a slight  mismatch between the OP profiles in the slab and in the wetting film
produce
deviations in amplitude of the scaling functions
comparable
to the   ones in panel (a) of each figure.
In addition, these
deviations
might be caused by
the difference between the
thermodynamic paths
taken
in the two panels.
In Fig.~\ref{SLAB_SCALING_super}(b)
the dashed
section
of the blue curve
(with $t>0$)
shows
that
part,
for which the
scaling hypothesis
breaks down.
This occurs for
very small values of
$L_0$,
in particular above
the tricritical
end point,
where the
wetting
film thickness
is
small, .
This is in line with the general rule that universal scaling functions
only hold in the scaling limit $L_0\gg a$.



\section{Summary and conclusions}

By using mean field theory,
layering transitions, wetting films, and tricritical Casimir forces (TCFs)
in $\Ht$ -$\Hf$ mixtures
have been studied
within the vectorized
Blume--Emery--Griffiths
model 
on
a semi-infinite, simple
cubic
lattice.
In the bulk, the model reduces to the one studied in Ref.~\cite{bulk-phase-diagram}.
For vanishing
coupling constant
$J_s$, which facilitates superfluid transitions,
the bulk phase diagram corresponds to
that of classical
binary liquid
mixtures
(Figs.~\ref{different_topologies}(a),~\ref{binary_fluid_schematic}).
We have identified
those
values of $J_s$ (see Fig.~\ref{topology_condition}), for which  the bulk phase diagram resembles
that of
actual $\Ht$ -$\Hf$ mixtures (Figs.~\ref{different_topologies}(b)~and~(c) and
Fig.~\ref{3d_no_path}).

The present model includes
short--ranged
surface fields
$f_+$ and
$f_-$
coupled to the sum and
to the
difference of the number densities 
of $\Ht$ and $\Hf$ atoms, respectively,
which
allows for
the occurrence of wetting
phenomena
and can control
the preference of
the surfaces
for the species.
The effect of the surface fields on wetting films  has been
studied for $J_s=0$.
Depending on the
values
of $f_+$ and $f_-$,
in the vapor phase very close to liquid-vapor coexistence,
the model exhibits
incomplete
or complete wetting
(Figs.~\ref{small_f_plus}-\ref{changing-offset-normal}).
Due to the  lattice character of the present  model,  we observe also
first-order layering transitions
(Figs.~\ref{layering}~and~\ref{film_thickness_Normal_case}).

For suitable values of the surface fields and for the coupling constants, which
determine the bulk phase diagram of the $\Ht$ -$\Hf$ mixtures,
we have been able to reproduce
qualitatively
the experimental results (see Fig.~\ref{Indirect-Fig}) for the thickness of $\Ht$ -$\Hf$
wetting films  near the tricritical end point~\cite{indirect}.
Although the measurements in Refs.~\cite{indirect} have been performed in the regime of complete wetting, due to gravity
the thickness of the
wetting films remained finite.
In the present study this is achieved
by applying
an offset to the experimental thermodynamic paths
(Fig.~\ref{path_he43}) and
shifting  them into the vapor phase so that the resulting wetting films remain finite
(Figs.~\ref{3d_no_path}~and~\ref{path_on_offset_plane}).
Within the present mean field approach
the order parameter profiles at a given thermodynamic state
provide all
equilibrium properties of the wetting films
(Fig.~\ref{profiles_at_three-T}).
The closer the system  to liquid-vapor coexistence is,
the thicker the wetting films are (Fig.~\ref{changing_offset_X3_constant}).
Depending on the thermodynamic state, the wetting films can be superfluid.
For the bulk phase corresponding to the normal fluid,
the onset of superfluidity occurs
by crossing a line of
continuous surface transitions (Fig.~\ref{bulk-modify}).

Taking  thermodynamic paths
(Fig.~\ref{bulk-modify})
equivalent to the experimental ones taken in Ref.~\cite{indirect},
we have been able to reproduce
qualitatively
the experimental results
for the variation of the film thickness upon approaching  the
tricritical end point.
Since the tricritical end point
lies between the wetting temperature
and the critical point of the
\REDCOLOR{liquid-vapor}
phase transitions,
there is
a pronounced change in the
thickness of the wetting film due to repulsive TCFs
(Figs.~\ref{thickness_all_T}~,~\ref{Xtce_different_offsets},~\ref{film_thickness_N},~and~\ref{film_thickness_S}).
The repulsive nature of
the
TCF is due to the
effectively
non--symmetric boundary conditions
for
the superfluid OP.
The non--symmetric boundary conditions arise due to
the formation of a $\Hf$-rich layer near the solid--liquid interface, which
can become superfluid even at temperatures above the $\lambda$-transition;
at the liquid--vapor interface such a superfluid layer does
not form because the $\Hf$ concentration is too low there.
This leads to $(+,O)$ boundary conditions.
Such boundary conditions hold below the line
\tcr{$T_{\text{s}}(\mathcal{C}_3)$}
of surface transitions
(blue curve in Fig.~\ref{bulk-modify})
up to the special point $s^*$
(i.e., for $\mathcal{C}_3>\mathcal{C}_3^{s^*}$).
Like
\tcr{the experiment data,}
upon decreasing the temperature
along the thermodynamic paths
at fixed $\mathcal{C}_3$
in the region
$\mathcal{C}_3^{s^*}<\mathcal{C}_3<\mathcal{C}_3^{\text{tce}}$,
in addition to the repulsive TCFs
close
\tcr{to}
tce
the wetting films are also influenced by the repulsive
critical Casimir forces (CCFs) close to the $\lambda$-line
\tcr{$T_{\lambda}(\mathcal{C}_3)$}
(red line in Fig.~\ref{bulk-modify}).
This gives rise to the formation of a
\tcr{shoulderlike}
curve
in Figs.~\ref{film_thickness_S} and \ref{Indirect-Fig}(b)
between the tricritical end point
and the $\lambda$-transition temperature.
For
$\mathcal{C}_3<\mathcal{C}_3^{s^*}$
the wetting film resembles that of pure $\Hf$, for which
the superfluid order parameter vanishes
\tcr{both}
at the solid substrate and at the liquid--vapor interface.
Such symmetric $(O,O)$
boundary conditions lead to
\tcr{an}
attractive
CCF, which results in the decrease of the wetting film thickness
close to the $\lambda$-transition temperature
\tcr{$T_{\lambda}(\mathcal{C}_3)$}
(see
the dip in Fig.~\ref{Indirect-Fig}(c)).
However,
\tcr{because}
the attractive CCF
\tcr{due to $(O,O)$ BC is generated by fluctuations only~\cite{1886}
it cannot
be captured}
within the present mean field approach.

Using the
various
contributions to the total free energy, 
one can calculate
the
TCFs and their scaling
function
by extracting the excess free energy
from the total free energy
(Figs.~\ref{SLAB_SCALING_normal}(a),~\ref{SLAB_SCALING_super}(a),~and~\ref{SLAB_SCALING_different_offsets}(a)).
We have adapted the slab  approximation for the wetting films to the present
system
and have calculated
the
corresponding slab
scaling function of the TCF
(Figs.~\ref{SLAB_SCALING_normal}(b),~\ref{SLAB_SCALING_super}(b),~and~\ref{SLAB_SCALING_different_offsets}(b)).
We have found  that  the slab
approximation,
with fixed surface fields
at
the second wall mimicking
the emerging
liquid--vapor
interface,
captures
rather well the qualitative
behavior of the scaling functions
inferred from the wetting film thickness
(see the comparison between the panels (a) and (b) in
Figs.~\ref{SLAB_SCALING_normal}-\ref{SLAB_SCALING_different_offsets}).

\begin{figure}
\includegraphics[width=120mm]{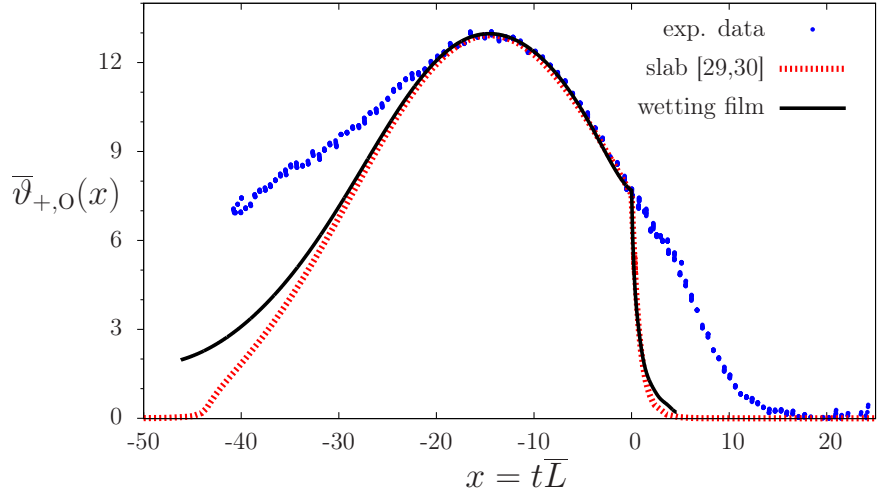}
\centering
\centering
\caption{\linespread{0.8}
\footnotesize{Adjusted
scaling functions (see the main text)
obtained
for the slab geometry
as in Refs.~\cite{maciole-dietrich-2006,maciolek-gambassi-dietrich}
and
inferred from
the wetting
films
compared with the
corresponding experimental curve~\cite{indirect}.
All data correspond to the
tricritical
concentration
of $\Ht$.
$\overline{L}$ is the film thickness
of the wetting films,
whereas in Refs.~\cite{maciole-dietrich-2006,maciolek-gambassi-dietrich}
it denotes the width of the slab.
The reduced temperature
$t=(T-T_\text{tc})/T_\text{tc}$ is
relative to
tricritical point
in Refs.~\cite{maciole-dietrich-2006,maciolek-gambassi-dietrich} and
relative to
the tricritical end point
for
the wetting film.
}
}
\label{compare}
\end{figure}

We conclude by comparing  the scaling function
inferred from
the wetting film
thickness and
the one calculated
within
the
slab geometry
as
in Refs.~\cite{maciole-dietrich-2006,maciolek-gambassi-dietrich} 
with the experimental data~\cite{indirect},
specifically
at the tricritical concentrations
$\mathcal{C}_3^\text{tce}$
of $\Ht$.
Figure~\ref{compare} illustrates this comparison.
$\overline{L}$ refers to the
wetting
film thickness measured
in the experimental data or
calculated within the present model.
In Refs.~\cite{maciole-dietrich-2006,maciolek-gambassi-dietrich}
$\overline{L}$ refers to the slab width.
In the reduced temperature
$t=(T-T_\text{tc})/T_\text{tc}$,
\tcr{$T_\text{tc}$}
refers to the temperature
of the tricritical end point both in the present calculation and in the experimental
studies, whereas it denotes the tricritical temperature
in Refs.~\cite{maciole-dietrich-2006,maciolek-gambassi-dietrich}.
The
theoretical scaling functions are rescaled such that their values
at $t=0$ match the experimental one. Moreover,
the scaling variable $x=t\overline{L}$
for
the theoretical results is multiplied by a
suitable
factor such that the
positions
of the maxima of
the
theoretical
curves match the experimental one.
This factor is $b_{\text{th}}\simeq23.1$ for the wetting film, whereas
for the slab geometry it is $b_\text{th}^{\text{VBEG}}\simeq15.38$.
The resulting adjusted scaling
functions  $\overline{\vartheta}_{+,O}(x)$ agree with each other 
and reproduce rather well the experimental
data, especially near the maximum. In contrast,
if
these two adjustments
of the scaling function
is enforced for the one
obtained within the slab approximation inferred from the wetting films
(i.e., the black curve in panel (b) of Fig.~\ref{SLAB_SCALING_super}),
there is no satisfactory
agreement with the experimental data
as a whole (this
adjusted
scaling function is not shown in Fig.~\ref{compare}).

The present model
lends itself to
further investigations based on Monte Carlo simulations.
They would capture the effects of fluctuations beyond the present mean field
theory. Since the upper critical dimension for tricritical phenomena
is
$d^*=3$,
this would shed additional light on the reliability of the present mean field analysis.
Moreover, in view of the ubiquity of van der Waals interactions it will be
rewarding to extend the present model by incorporating
long--ranged surface fields.

\section{Acknowledgments}
\REDCOLOR{N. Farahmand Bafi would like to thank
Dr. Markus Bier and Dr. Piotr Nowakowski
for fruitful discussions.}

\appendix

\section{Mean field approximation for the lattice model}

In this appendix
we present the details of the calculations
outlined in
Subsec.~\ref{subsec:mfa}.
The
starting point is the Hamiltonian in Eq.~(\ref{hamiltonian-summation}). 
According to the variation principle, the
equilibrium free energy $F$ obeys the inequality~\cite{chaikin}
\begin{equation}
\label{var-principle}
F \leq \phi=\hat{\text{T}}\text{r}( \rho \mathcal{H}) + (1/\beta) \hat{\text{T}}\text{r}(\rho \ln \rho)\text{,}
\end{equation}
where $\rho$ is any trial density matrix
fulfilling $\hat{\text{T}}\text{r}(\rho)=1$,
with respect to which $\phi$ on the rhs of
Eq.~(\ref{var-principle}) has to
be minimized in order to obtain the best approximation for $F$.
\begin{equation}
\label{trace-formula}
\hat{\text{T}}\text{r}=\sum_{s_1=\pm 1,0}\int_0^{2\pi}\text{d}\Theta_1 \cdot... \cdot\sum_{s_{L\mathcal{N}}=\pm 1,0}\int_0^{2\pi}\text{d}\Theta_{L\mathcal{N}}
\end{equation}
denotes the trace
and $\beta=1/T$ where $T$ is the temperature times $k_\text{B}$.
Within mean field theory,
the total density matrix
of the system
factorizes as
\begin{equation}
\label{RHO-first}
 \rho=\prod_{i=1}^{L\mathcal{N}} \rho _{i}=\prod_{l=0}^{L-1}\prod_{v_l=1}^{\mathcal{N}}\rho _{(l,v_l)}
\end{equation}
with 
\begin{equation}
\text{Tr}\rho _{(l,v_l)}=\sum_{s_{(l,v_l)}=\pm 1,0}\int_0^{2\pi}\text{d}\Theta_{(l,v_l)} \rho _{(l,v_l)}(s_{(l,v_l)},\Theta_{(l,v_l)})=1,
\end{equation}
where $l$ labels the
$L$
layers, $v_l$ denotes the lattice sites
within the $l^{\text{th}}$ layer, and $\rho _{(l,{v_l})}$
denotes the density matrix of
lattice site $v_l$ within the layer $l$.
(Note that $\hat{\text{T}}\text{r}$ denotes the trace over all
degrees of freedom,
whereas $\text{Tr}$
refers to the trace over
the degrees of freedom at
a single lattice site.)

By applying
mean field approximation
to the sites within each
layer, $\rho _{(l,{v_l})}$ is taken to be independent
of $v_l$. Accordingly,
Eq.~(\ref{RHO-first}) renders

\begin{equation}
\label{RHO}
 \rho=\prod_{l=0}^{L-1}\rho _{l}^{\mathcal{N}},
\end{equation}
with
\begin{equation}
\text{Tr}\rho _{l}=\sum_{s_{l}=\pm 1,0}\int_0^{2\pi}\text{d}\Theta_{l} \rho _{l}(s_{l},\Theta_{l})=1,
\end{equation}
where $\rho_{l}\equiv\rho_{(l,v_l)}$ indicates
the density matrix for a single site in the $l^{\text{th}}$
layer;
$s_{l}\equiv s_{(l,v_l)}$ and $\Theta_{l}\equiv \Theta_{(l,v_l)}$
denote the occupation variable
and the angle
for a single site within this layer,
respectively,
independent of $v_l$. (Note that due to the definitions in
Eq.~(\ref{coupling}), one has $q_l\equiv q_{(l,v_l)}$ and $p_l\equiv p_{(l,v_l)}$.)
The summations in Eq.~(\ref{hamiltonian-summation}) can be written as
\begin{equation}
\label{breaking-sum}
\sum_{i=1}^{L\mathcal{N}}=\sum_{l=0}^{L-1}\sum_{{v_l}=1}^{\mathcal{N}}=\mathcal{N}\sum_{l=0}^{L-1}
\end{equation}
and
\begin{equation}
\begin{split}
\sum_{<i,j>}& =\frac{1}{2} \sum_{i=1}\{ \sum_{j\in \text{n.n.$(l)$}}+\sum_{j\in \text{n.n.$(l+1)$}}+\sum_{j\in \text{n.n.$(l-1)$}}(1-\delta_{l,0}) \}\\ 
	    &=\frac{\mathcal{N}}{2} \sum_{l=0}^{L-1}\{ 4+\sum_{j\in \text{n.n.$(l+1)$}}+\sum_{j\in \text{n.n.$(l-1)$}}(1-\delta_{l,0}) \}
\end{split}
\end{equation}
where n.n.$(l)$, n.n.$(l+1)$, and n.n.$(l-1)$ denote the nearest neighbors in
the layers $l$, $l+1$, and
$l-1$, respectively.
The factor $1/2$
prevents
double counting and
the factor $(1-\delta_{l,0})$ appears due to the fact that
layer $l=0$
next
to the surface
does not have a neighboring layer at $l=-1$.
Since the lattice sites within each layer are
equivalent one has
$\sum_{j\in \text{n.n.$(l)$}}=4$.

By
using Eq.~(\ref{hamiltonian-summation}) together with the above considerations,
Eq.~(\ref{var-principle}) renders
\begin{widetext}
\begin{equation}
\label{phi1}
\begin{split}
 \phi= & -\frac{K\mathcal{N}}{2}\sum_{l=0}^{L-1} \langle s_{l}\rangle(4\langle s_{l}\rangle+\langle s_{l+1}\rangle+\langle s_{l-1}\rangle(1-\delta_{l,0}))\\
       & -\frac{J\mathcal{N}}{2}\sum_{l=0}^{L-1} \langle q_{l}\rangle(4\langle q_{l}\rangle+\langle q_{l+1}\rangle+\langle q_{l-1}\rangle(1-\delta_{l,0}))\\
       & -\frac{C\mathcal{N}}{2}\sum_{l=0}^{L-1} \langle s_{l}\rangle(4\langle q_{l}\rangle+\langle q_{l+1}\rangle+\langle q_{l-1}\rangle(1-\delta_{l,0}))\\
       & -\frac{C\mathcal{N}}{2}\sum_{l=0}^{L-1} \langle q_{l}\rangle(4\langle s_{l}\rangle+\langle s_{l+1}\rangle+\langle s_{l-1}\rangle(1-\delta_{l,0}))\\
       & -\frac{J_s\mathcal{N}}{2}\sum_{l=0}^{L-1}\langle p_{l} \cos \Theta_{l} \rangle(4\langle p_{l} \cos \Theta_{l}\rangle+\langle p_{l+1} \cos \Theta_{l+1}\rangle+\langle p_{l-1} \cos \Theta_{l-1}\rangle(1-\delta_{l,0}))\\
       & -\frac{J_s\mathcal{N}}{2}\sum_{l=0}^{L-1}\langle p_{l} \sin \Theta_{l} \rangle(4\langle p_{l} \sin \Theta_{l}\rangle+\langle p_{l+1} \sin \Theta_{l+1}\rangle+\langle p_{l-1} \sin \Theta_{l-1}\rangle(1-\delta_{l,0}))\\
       & -\mathcal{N}\mu_{-}\sum_{l=0}^{L-1}\langle s_{l}\rangle-\mathcal{N}\mu_{+}\sum_{l=0}^{L-1}\langle q_{l}\rangle\\
       & -\mathcal{N}\sum_{l=0}^{L-1}\langle f_{-}(l)s_{l}\rangle-\mathcal{N}\sum_{l=0}^{L-1}\langle f_{+}(l) q_{l}\rangle\\
       & +(1/ \beta)\langle \ln \prod_{l=0}^{L-1}\rho _{l}^{\mathcal{N}} \rangle,\\
\end{split}
\end{equation}
\end{widetext}
where $\langle...\rangle=\text{Tr}(\rho_l...)$ denotes the thermal average
taken with the trial density matrix $\rho_{l}$ associated with a single lattice site
in layer $l$.

The last term
in Eq.~(\ref{phi1})
can be written as
\begin{equation}
\label{product-to-sum}
(1/ \beta)\langle \ln \prod_{l=0}^{L-1}\rho _{l}^{\mathcal{N}} \rangle=(\mathcal{N}/ \beta)\langle  \sum_{l=0}^{L-1}\ln \rho _{l} \rangle.
\end{equation}

Minimizing
the variational function $\phi/\mathcal{N}$
with respect to $\rho_{l}$ renders
the best
normalized
functional form
of $\rho_{l}$
among the single--site, factorized density matrices.
Thus we determine
the functional derivative of $\phi/\mathcal{N}$
in Eq.~(\ref{phi1})
with respect to $\rho_{l}(s_l,\Theta_l)$
using
$\frac{\delta \rho_{l}(s_l,\Theta_l)}{\delta \rho_{l'}(s'_{l'},\Theta'_{l'})}=\delta_{l,l'} \delta(\Theta_l-\Theta'_{l'})\delta_{s_l,s'_{l'}}$,
and equate it to the Lagrange multiplier $\eta$
corresponding to the
constraint
$\text{Tr}(\rho_l)=1$:
\begin{widetext}
\begin{equation}
\label{derivative-of-phi}
\begin{split}
 \eta=\frac{\delta( \phi/\mathcal{N})}{\delta\rho_{l}(s_l,\Theta_l)} =& -K\{ s_l(4X_l+X_{l+1}+X_{l-1}(1-\delta_{l,0}))\} \\
						       & -J\{ q_l(4D_l+D_{l+1}+D_{l-1}(1-\delta_{l,0}))\} \\
						       & -C\{ s_l(4D_l+D_{l+1}+D_{l-1}(1-\delta_{l,0}))\} \\
						       & -C\{ q_l(4X_l+X_{l+1}+X_{l-1}(1-\delta_{l,0}))\} \\
						       & -(\mu_{-}+f_{-}(l))s_l-(\mu_{+}+f_{+}(l))q_l\\
						       & -J_s\{ p_l\cos \Theta_l(4M_l^x+M_{l+1}^x+M_{l-1}^x(1-\delta_{l,0}))\} \\
						       & -J_s\{ p_l\sin \Theta_l(4M_l^y+M_{l+1}^y+M_{l-1}^y(1-\delta_{l,0}))\} \\
						       & + (1/ \beta) (1+\ln \rho_l)
\end{split}
\end{equation}
\end{widetext}
where we have defined the following order parameters (OPs) 
\begin{equation}
\label{definition of order parameters_APPENDIX}
 \begin{split}
   & X_l:=\langle s_{l}\rangle\text{,}\\
   & D_l:=\langle q_{l}\rangle\text{,}\\
   & M^{x}_l:=\langle p_{l}\cos\Theta_{l}\rangle\text{,}\\
   & M^{y}_l:=\langle p_{l}\sin\Theta_{l}\rangle\text{.}  
 \end{split}
\end{equation}
Equation (\ref{derivative-of-phi}) can be solved for $\rho_{l}(s_l,\Theta_l)$:
\begin{equation}
\rho_{l}=e^{\beta \eta -1-\beta h_{l}}\text{,}
\end{equation}
where
\begin{equation}
\label{hL}
\begin{split}
h_l=&-s_l\{ K(4X_l+X_{l+1}+X_{l-1}(1-\delta_{l,0}))+C(4D_l+D_{l+1}+D_{l-1}(1-\delta_{l,0}))+\mu_{-}+f_{-}(l)\}\\
    &-q_l\{ J(4D_l+D_{l+1}+D_{l-1}(1-\delta_{l,0}))+C(4X_l+X_{l+1}+X_{l-1}(1-\delta_{l,0}))+\mu_{+}+f_{+}(l)\}\\
    &-p_l\cos\Theta_l\{ J_s(4M^x_l+M^x_{l+1}+M^x_{l-1}(1-\delta_{l,0}))\}\\
    &-p_l\sin\Theta_l\{ J_s(4M^y_l+M^y_{l+1}+M^y_{l-1}(1-\delta_{l,0}))\}\\
\end{split}
\end{equation}
is the effective
single-site Hamiltonian for a lattice site in the $l^{\text{th}}$ layer.
\\
The normalization $\text{Tr}(\rho_l)=1$ yields
\begin{equation}
e^{-\beta \eta +1}=\text{Tr}(e^{-\beta h_{l}})
\end{equation}
so that
\begin{equation}
\label{rho}
\rho_{l}=\frac{e^{-\beta h_{l}}}{\text{Tr}(e^{-\beta h_{l}})}\text{,}
\end{equation}
where $h_{l}$ is given by Eq.~(\ref{hL}).

Within
the expression for $h_l$
given in Eq.~(\ref{hL})
one has
\begin{equation}
\label{trace-value}
\begin{split}
\text{Tr}e^{-\beta h_{l}}=& 1+W_l(X_l,D_l;\mu_{-},\mu_{+},f_+(l),f_-(l),T)\\
		    & +R_l(X_l,D_l;\mu_{-},\mu_{+},f_+(l),f_-(l),T) I_0(\beta J_s \tilde{M_l}),
\end{split}
\end{equation}
where $I_{0}$ and $I_{1}$ are modified Bessel functions
(see
Subsec.~9.6
in Ref.~\cite{abramowitz}) and 
\begin{equation}
\tilde{M_l}=\sqrt{(M^x_{l-1}(1-\delta_{l,0})+4 M^x_{l}+M^x_{l+1})^2+(M^y_{l-1}(1-\delta_{l,0})+4 M^y_{l}+M^y_{l+1})^2}.
\end{equation}

The functions
$W(X_l,D_l;\mu_{-},\mu_{+},f_+(l),f_-(l),T)$ and  $R(X_l,D_l;\mu_{-},\mu_{+},f_+(l),f_-(l),T)$ are given by
\begin{equation}
\label{EqW_APPENDIX}
\begin{split}
W_l(X_l,D_l;\mu_{-},\mu_{+},f_+(l),f_-(l),T)=\exp\beta\{&(J-C) (D_{l-1}(1-\delta_{l,0})+4 D_{l}+D_{l+1})\\
						    & +(C-K)(X_{l-1}(1-\delta_{l,0})+4 X_{l}+X_{l+1})\\
						    &+ \mu_{+}+f_{+}(l)- \mu_{-}-f_{-}(l)\}
\end{split}
\end{equation}
and
\begin{equation}
\label{EqR_APPENDIX}
\begin{split}
R_l(X_l,D_l;\mu_{-},\mu_{+},f_+(l),f_-(l),T)=\exp\beta\{&(J+C) (D_{l-1}(1-\delta_{l,0})+4 D_{l}+D_{l+1})\\
						    &+(C+K)(X_{l-1}(1-\delta_{l,0})+4 X_{l}+X_{l+1})\\
						    &+ \mu_{+}+f_{+}(l)+\mu_{-}+f_{-}(l)\}.
\end{split}
\end{equation}

Using the definitions in
Eq.~(\ref{definition of order parameters_APPENDIX})
the OPs are given by four coupled self-consistent equations:
\begin{equation}
\label{eqX_APPENDIX}
X_l=\frac{-W_l+R_l I_0(\beta J_s \tilde{M_l})}{1+W_l+R_l I_0(\beta J_s\tilde{M_l})}
\end{equation}
and
\begin{equation}
\label{eqD_APPENDIX}
D_l=\frac{W_l+R_l I_0(\beta J_s \tilde{M_l})}{1+W_l+R_l I_0(\beta J_s \tilde{M_l})}\text{;}
\end{equation}
$M_l^x$ and $M_l^y$ are given by
\begin{equation}
\label{eqMx_APPENDIX}
M_l^x=\frac{(1-\delta_{l,0})M^x_{l-1}+4 M^x_{l}+M^x_{l+1}}{\tilde{M_l}}\frac{R_l I_1(\beta J_s \tilde{M_l})}{1+W_l+R_l I_0(\beta J_s \tilde{M_l})}
\end{equation}
and
\begin{equation}
\label{eqMy_APPENDIX}
M_l^y=\frac{(1-\delta_{l,0})M^y_{l-1}+4 M^y_{l}+M^y_{l+1}}{\tilde{M_l}}\frac{R_l I_1(\beta J_s \tilde{M_l})}{1+W_l+R_l I_0(\beta J_s \tilde{M_l})}
\end{equation}
so that
\begin{equation}
\label{eqM_APPENDIX}
M_l:=\sqrt{(M_l^x)^2+(M_l^y)^2}=\frac{R_l I_1(\beta J_s \tilde{M_l})}{1+W_l+R_l I_0(\beta J_s \tilde{M_l})}.
\end{equation}
Since $(M_l^x, M_l^y)$ and $\tilde{M_l}$ are invariant under rotation around the
$z$-axis,
it is sufficient
to consider only one of the two
components. We choose a rotation such that $M_l^y=0$ and $M_l^x>0$.
With this choice one has
\begin{equation}
\tilde{M_l}=(1-\delta_{l,0})M^x_{l-1}+4 M^x_{l}+M^x_{l+1}
\end{equation}
and
\begin{equation}
\label{eqM-easy_APPENDIX}
M_l=\sqrt{(M_l^x)^2+(M_l^y)^2}=M_l^x=\frac{R_l I_1(\beta J_s \tilde{M_l})}{1+W_l+R_l I_0(\beta J_s \tilde{M_l})}.
\end{equation}
\tcb{In order to determine
the equilibrium free energy
given in Eq.~(\ref{phi1})
we first
rearrange the term
$(1/ \beta)\langle \ln \prod_{l=0}^{L-1}\rho _{l}^{\mathcal{N}} \rangle$
(see also Eq.~(\ref{product-to-sum}):}
\begin{equation}
\begin{split}
\label{simplifying-trace}
&(1/ \beta)\langle \ln \prod_{l=0}^{L-1}\rho _{l}^{\mathcal{N}} \rangle=(\mathcal{N}/ \beta)\langle  \sum_{l=0}^{L-1}\ln \rho _{l} \rangle=(\mathcal{N}/ \beta)\sum_{l=0}^{L-1}\langle \ln \frac{e^{-\beta h_{l}}}{\text{Tr} e^{-\beta h_{l}}}\rangle\\
&\quad\quad\quad\quad\quad\quad\quad\quad\quad=-\mathcal{N}\sum_{l=0}^{L-1}\langle h_l \rangle-(\mathcal{N}/\beta)\sum_{l=0}^{L-1}\langle \ln \text{Tr} e^{-\beta h_{l}}\rangle\text{,}
\end{split}
\end{equation}
where in the last
step, using Eqs.~(\ref{trace-value})~and~(\ref{eqD_APPENDIX}),
we can write $\text{Tr} e^{-\beta h_{l}}=(1-D_l)^{-1}$.

Inserting $\rho_l$
into
Eq.~(\ref{phi1}) with
the choice
$M_l^y=0$ and $M_l^x>0$
and
taking into account Eq.~(\ref{simplifying-trace}) one obtains
the following
mean field
expression for the equilibrium free energy:
\begin{equation}
\label{free-energy-simple_APPENDIX}
\begin{split}
 \phi/\mathcal{N}=& \sum_{l=0}^{L-1} \Big [\frac{K}{2} X_l(4X_l+X_{l+1}+X_{l-1}(1-\delta_{l,0}))\\
		  & \quad \quad +\frac{J}{2} D_l(4D_l+D_{l+1}+D_{l-1}(1-\delta_{l,0}))\\
		  & \quad \quad +\frac{C}{2} X_l(4D_l+D_{l+1}+D_{l-1}(1-\delta_{l,0}))\\
		  & \quad \quad +\frac{C}{2} D_l(4X_l+X_{l+1}+X_{l-1}(1-\delta_{l,0}))\\
		  & \quad \quad +\frac{J_s}{2} M_l^x(4M_l^x+M^x_{l+1}+M^x_{l-1}(1-\delta_{l,0}))\\
		  & \quad \quad +(1/\beta)\ln (1-D_l) \Big ].
\end{split}
\end{equation}
Note that in
the general case
(i.e., for both $M_l^y$ and $M_l^x$ being nonzero) the
contribution
$\frac{J_s}{2} M_l^y(4M_l^y+M^y_{l+1}+M^y_{l-1}(1-\delta_{l,0}))$ has to be added
to the rhs of Eq.~(\ref{free-energy-simple_APPENDIX}).

In order to obtain
the functional form of the expressions
\tcr{for}
the
chemical
potentials,
first Eqs.~(\ref{eqX_APPENDIX})~and~(\ref{eqD_APPENDIX})
have to be solved for $W_l$ and $R_l$.
Then, by
comparing these solutions with the definitions of $W_l$
and
$R_l$ as in
Eqs.~(\ref{EqW_APPENDIX})~and~(\ref{EqR_APPENDIX}), one finds
\begin{equation}
\label{deltaPLUS_APPENDIX}
 \begin{split}
     \mu_{+}=& \frac{T}{2}\ln (D_l^2-X_l^2)-T\ln2-T\ln(1-D_l)-\frac{T}{2}\ln(I_0(\beta J_s \tilde{M_l}))\\
                                  & -J(D_{l-1}(1-\delta_{l,0})+4D_{l}+D_{l+1})-C(X_{l-1}(1-\delta_{l,0})+4X_{l}+X_{l+1})-f_{+}(l)
 \end{split}
\end{equation}
and
\begin{equation}
\label{deltaMINUS_APPENDIX}
\begin{split}
\mu_{-}=& \frac{T}{2}\ln\frac{D_l+X_l}{D_l-X_l}-\frac{T}{2}\ln(I_0(\beta J_s \tilde{M_l}))\\
			     & -C(D_{l-1}(1-\delta_{l,0})+4D_{l}+D_{l+1})-K(X_{l-1}(1-\delta_{l,0})+4X_{l}+X_{l+1})-f_{-}(l).
\end{split}
\end{equation}

Finally,
one can
implicitly
express the magnetization $M_l$ in terms of $X_l$
and $D_l$ by
using
Eqs.~(\ref{eqX_APPENDIX}),~(\ref{eqD_APPENDIX}),~and~(\ref{eqM_APPENDIX}):
\begin{equation}
\label{mEQUILIBRIUM_APPENDIX}
\frac{X_l+D_l}{2}=\frac{M_l I_0(\beta J_s \tilde{M_l})}{I_1(\beta J_s \tilde{M_l})}.
\end{equation}
%

%

\clearpage



\begin{thebibliography}{10}




\bibitem{generalphasediagram}
M.~Krech and S.~Dietrich,
\newblock Phys. Rev. A {\bf 46}, 1922 (1992).

\bibitem{Chanhe4}
R.~Garcia and M.~H.~W. Chan,
\newblock Phys. Rev. Lett. {\bf 83}, 1187 (1999).

\bibitem{indirect}
R.~Garcia and M.~H.~W. Chan,
\newblock Phys. Rev. Lett. {\bf 88}, 086101 (2002).

\bibitem{Ganshin}
A.~Ganshin, S.~Scheidemantel, R.~Garcia, and M.~H.~W. Chan,
\newblock Phys. Rev. Lett. {\bf 97}, 075301 (2006).

\bibitem{Fukuto}
M.~Fukuto, Y.~F. Yano, and P.~S. Pershan,
\newblock Phys. Rev. Lett. {\bf 94}, 135702 (2005).

\bibitem{Rafai}
S.~Rafa\"\i, D.~Bonn, and M.~J.,
\newblock Physica {\bf 386}, 31 (2007).

\bibitem{Mukhopadhyay}
A.~Mukhopadhyay and B.~M. Law,
\newblock Phys. Rev. Lett. {\bf 83}, 772 (1999).

\bibitem{Mukhopadhyay2}
A.~Mukhopadhyay and B.~M. Law,
\newblock Phys. Rev. E {\bf 62}, 5201 (2000).

\bibitem{Fischer-deGennes-1978}
M.~E. Fisher and P.~G. de~Gennes,
\newblock C. R. Seances Acad. Sci. Paris Ser. B {\bf 287}, 207 (1978).

\bibitem{FSS}
K.~Binder,
\newblock in {\em Phase Transitions and Critical Phenomena}, edited by C.~Domb
  and J.~L. Lebowitz (Academic, London, 1983), Vol.~8, p. 149.


\bibitem{FSS2}
V.~Privman,
\newblock in {\em Finite Size Scaling and Numerical Simulation of Statistical
  Systems}, edited by V.~Privman (World Scientific, Singapore, 1990), p.~1.


\bibitem{casimiroriginal}
H.~Casimir,
\newblock Proc. K. Ned. Akad. Wet {\bf 51}, 793 (1948).

\bibitem{krech}
M.~Krech,
\newblock {\em The Casimir effect in critical systems}
\newblock (World Scientific, Singapore, 1994).

\bibitem{nightingale}
M.~P. Nightingale and J.~O. Indekeu,
\newblock Phys. Rev. Lett. {\bf 54}, 1824 (1985).

\bibitem{prl66}
M.~Krech and S.~Dietrich,
\newblock Phys. Rev. Lett. {\bf 66}, 345 (1991).

\bibitem{1886}
M.~Krech and S.~Dietrich,
\newblock Phys. Rev. A {\bf 46}, 1886 (1992).

\bibitem{Diehl}
H.~W. Diehl,
\newblock in {\em Phase Transitions and Critical Phenomena}, edited by C.~Domb
  and J.~L. Lebowitz (Academic, London, 1986), Vol.~10, p.~ 75.


\bibitem{danchev}
J.~G. Brankov, D.~Danchev, and N.~Tonchev,
\newblock {\em Theory of critical phenomena in finite--size systems}
(World Scientific, Singapore, 2000).


\bibitem{Zandi-Rudnick-Kardar-2004}
R.~Zandi, J.~Rudnick, and M.~Kardar,
\newblock Phys. Rev. Lett. {\bf 93}, 155302 (2004).
\REDCOLOR{
\bibitem{Maciolek-Gambassi-Dietrich-2007}
A.~Macio{\l}ek, A.~Gambassi, and S.~Dietrich,
\newblock Phys. Rev. E {\bf 76}, 031124 (2007).
}
\bibitem{Zandi-Shackel-Rudnick-Kardar-Chayes-2007}
R.~Zandi, A.~Shackell, J.~Rudnick, M.~Kardar, and L.~P. Chayes,
\newblock Phys. Rev. E {\bf 76}, 030601(R) (2007).

\bibitem{Vasilyev-Gambassi-Maciolek-Dietrich-2007}
O.~Vasilyev, A.~Gambassi, A.~Macio{\l}ek, and S.~Dietrich,
\newblock EPL {\bf 80}, 60009 (2007).

\bibitem{Vasilyev-Gambassi-Maciolek-Dietrich-2009}
O.~Vasilyev, A.~Gambassi, A.~Macio{\l}ek, and S.~Dietrich,
\newblock Phys. Rev. E {\bf 79}, 041142 (2009).

\bibitem{Dantchev-Krech-2004}
D.~Dantchev and M.~Krech,
\newblock Phys. Rev. E {\bf 69}, 046119 (2004).

\bibitem{Hutch-2007}
A.~Hucht,
\newblock Phys. Rev. Lett. {\bf 99}, 185301 (2007).

\bibitem{hasenbusch-2009}
M.~Hasenbusch,
\newblock J. Stat. Mech.: Theory and Experiment {\bf 2009},
  P07031 (2009).

\bibitem{hasenbusch-2010}
M.~Hasenbusch,
\newblock Phys. Rev. B {\bf 81}, 165412 (2010).

\bibitem{Romagnan-et:1978}
J.~P. Romagnan, J.~P. Laheurte, J.~C. Noiray, and W.~F. Saam,
\newblock J. Low Temp. Phys. {\bf 30}, 425 (1978).

\bibitem{maciole-dietrich-2006}
A.~Macio{\l}ek and S.~Dietrich,
\newblock Europhys. Lett. {\bf 74}, 22 (2006).

\bibitem{maciolek-gambassi-dietrich}
A.~Macio{\l}ek, A.~Gambassi, and S.~Dietrich,
\newblock Phys. Rev. E {\bf 76}, 031124 (2007).

\bibitem{beg}
M.~Blume, V.~J. Emery, and R.~B. Griffiths,
\newblock Phys. Rev. A {\bf 4}, 1071 (1971).

\bibitem{2dvbeg2}
A.~N. Berker and D.~R. Nelson,
\newblock Phys. Rev. B {\bf 19}, 2488 (1979).

\bibitem{2dvbeg1}
J.~L. Cardy and D.~J. Scalapino,
\newblock Phys. Rev. B {\bf 19}, 1428 (1979).

\bibitem{bulk-phase-diagram}
N.~Farahmand~Bafi, A.~Macio{\l}ek, and S.~Dietrich,
\newblock Phys. Rev. E {\bf 91}, 022138 (2015).
%

\bibitem{gsl-cite}
M.~Galassi, J.~Davies, J.~Theiler, B.~Gough, G.~Jungman, P.~Alken, M.~Booth, and F.~Rossi, 
\newblock {\em GNU Scientific Library Reference Manual},
\newblock Network {T}heory {L}td., 2009;
\newblock library available online at \url{http://www.gnu.org/software/gsl/}.

%
\bibitem{vbeg3}
A.~Macio{\l}ek, M.~Krech, and S.~Dietrich,
\newblock Phys. Rev. E {\bf 69}, 036117 (2004).

\bibitem{bell}
G.~M. Bell and D.~A. Lavis,
\newblock {\em Statistical mechanics of lattice models,Vol.1},
\newblock (Springer, Chichester, 1989).
\bibitem{dietrich-wetting}
S.~Dietrich,
\newblock in {\em Phase Transitions and Critical Phenomena},
edited by C.~Domb and J.~L.~Lebowitz
(Academic, London, 1988), Vol.~12, p.~1.

\bibitem{dietrich-surface-films}
S.~Dietrich,
\newblock in {\em Phase Transitions in Surface Films 2}, proceedings of the NATO ASI (Series B) held in Erice, Italy, 19-30 June 1990,
edited by H. Taub, G. Torzo, H. J. Lauter, and S. C. Fain
(Plenum, New York, 1991), Vol.~B 267, p.~391.
  
  
\bibitem{Evans:1990}
R.~Evans,
\newblock J. Phys.: Condens. Matter {\bf 2}, 8989 (1990).

\bibitem{Krech:1997}
M.~Krech,
\newblock Phys. Rev. E {\bf 56}, 1642 (1997).

\bibitem{Gambassi:2009}
A.~Gambassi,
\newblock J. Phys.: Conf. Series {\bf 161}, 012037 (2009).


\bibitem{dietrich-napiorkowski-1991}
S.~Dietrich and M.~Napi\'orkowski,
\newblock Phys. Rev. A {\bf 43}, 1861 (1991).
\tcr{
\bibitem{binder-landau-mueller-2002}
K.~Binder, D.~Landau, and M.~M\"uller,
\newblock J. Stat. Phys. {\bf 110}, 1411 (2003).
}
\bibitem{riedel}
E.~K. Riedel,
\newblock Phys. Rev. Lett. {\bf 28}, 675 (1972).

\bibitem{sarbach}
I.~D. Lawrie and S.~Sarbach,
\newblock in {\em Phase Transitions and Critical Phenomena,}
edited by C. Domb and J. L. Lebowitz
(Academic, London, 1984), Vol.~9, p.~2.

\bibitem{chaikin}
P.~M. Chaikin and T.~Lubensky,
\newblock {\em Principles of condensed matter physics}
\newblock (Cambridge University press, Cambridge, 1995).


\bibitem{abramowitz}
M.~Abramowitz and I.~A. Stegun,
\newblock {\em \textnormal{eds.,} Handbook of mathematical functions}
\newblock (Dover, New York, 1972).





\end{thebibliography}

\end{document}